\documentclass[12pt]{article}
\usepackage{epsfig,array,graphicx,a4}
\newcommand{\hbo}{\hbox to 1 true cm {\hfill } } 
\newcommand{\tr}{\hbox{tr}}

\newcommand{\Id}{1}

\begin{document}  
\thispagestyle{empty}
\begin{flushright}
hep-lat/0412032 \\
December 21, 2004
\end{flushright} 
\vskip10mm
\begin{center}
{\LARGE
Center vortices and Dirac eigenmodes in 
\vskip2mm
SU(2) lattice gauge theory}
\vskip11mm
{\bf Jochen Gattnar$^a$, Christof Gattringer$^b$, 
Kurt Langfeld$^a$,\\ Hugo Reinhardt$^a$,
Andreas Sch\"afer$^b$, Stefan Solbrig$^b$ and T. Tok$^a$}
\vskip3mm
$^a$ Institut f\"ur Theoretische Physik, Universit\"at
T\"ubingen \\
D-72076 T\"ubingen, Germany
\vskip2mm
$^b$ Institut f\"ur Theoretische Physik, Universit\"at
Regensburg \\
D-93040 Regensburg, Germany 
\vskip20mm
\begin{abstract}  
We study the interplay between Dirac eigenmodes and center vortices in SU(2)
lattice gauge theory. In particular we focus on vortex-removed
configurations and compare them to an ensemble of configurations with 
random changes of the link variables. We show 
that removing the vortices destroys all zero modes and the near zero modes
are no longer coupled to topological structures. The Dirac
spectrum for vortex-removed configurations in many respects
resembles a free spectrum thus leading to a 
vanishing chiral condensate. Configurations 
with random changes leave the topological features of the Dirac eigensystem
intact. We finally show that smooth center vortex configurations
give rise to zero modes and topological near zero modes.  
\end{abstract}
\vskip10mm
{\sl To appear in Nuclear Physics B.}
\end{center}

\newpage

\setcounter{page}{1}


\section{Introduction and outline of results}

The QCD vacuum has a highly non-trivial structure. Its excitations give rise 
to the key features of QCD, namely confinement and chiral symmetry breaking. 
Over the years, for both features several mechanisms have been proposed.
For confinement the two most influential pictures are based on monopoles,
respectively center vortices, while for chiral symmetry breaking 
instanton-type excitations have played a major role. It is widely
expected that confinement and chiral symmetry breaking are linked through 
some unifying mechanism. This belief is supported by numerical results
indicating that 
at the QCD phase transition deconfinement and restoration of
chiral symmetry take place at the same critical temperature. However,
our understanding of such a unifying mechanism is still in its
infancy. 

Lattice QCD is an important tool for analyzing excitations of the 
QCD vacuum and is even necessary for formulating some of the mechanisms
such as the picture based on center vortices. Lattice simulations 
provide one with thermalized configurations of the gauge field as they appear 
in the path integral. These configurations can then be analyzed and one
can try to identify the excitations of the gluon field relevant for
different physical features. 

Such lattice investigations have provided strong evidence for the center 
vortex picture of confinement. Vortices obtained after center projection 
in the maximum center gauge \cite{DelDebbio:1996mh}, are physical 
(in the sense of the renormalization group)~\cite{Langfeld:1997jx}, 
and seem to 
constitute the relevant infrared degrees of freedom of Yang-Mills 
theory. Indeed, when center vortices are removed from the Yang-Mills
ensemble \cite{deForcrand:1999ms}, the confinement properties are lost
\cite{deForcrand:1999ms,Alexandrou:1999vx}.
The loss of confinement is indicated by the vanishing of the string tension 
and the change of the infrared behavior of the Green's functions in Landau 
gauge~\cite{Langfeld:2001cz,Gattnar:2004bf} and in Coulomb 
gauge~\cite{Greensite:2004ur}. 
The emergence of a string tension can be easily understood 
in the center vortex picture. In addition, center vortices provide also 
an appealing picture of the finite temperature deconfinement phase 
transition as a depercolation transition
in a 3-dimensional slice of the lattice universe, taken at a fixed spatial
coordinate~\cite{Langfeld:1998cz,Engelhardt:1999fd}. 
Moreover, center vortices reproduce not only 
the correct order of the phase transition but also 
its universality class~\cite{Langfeld:2003zi}.
On the other hand in the spatial 3-volume, there is no 
depercolation
transition and the center vortex ensemble correctly describes the increase of
the spatial string tension above the critical 
temperature~\cite{Gattnar:2000ke}. 

All these properties of center vortices detected on the lattice after center
projection, as well as the correct order of the deconfinement phase transition
are well reproduced in a random center vortex model for both 
SU(2)~\cite{Engelhardt:1999wr} and SU(3)~\cite{Engelhardt:2003wm}. 
Furthermore, the vortex percolation in the confinement
phase is consistent with the vanishing of the free energy of center vortices in
this phase as observed on the lattice~\cite{Kovacs:2000sy} 
and in the continuum~\cite{Lange:2003ti}. 

When center vortices are removed from the Yang-Mills ensemble, not only 
the string tension is gone but also the quark condensate vanishes and the 
topological charge is lost~\cite{deForcrand:1999ms,Alexandrou:1999vx}.
These findings suggest that center vortices could be responsible
for all infrared key features, confinement, chiral symmetry breaking and the
chiral anomaly. 

In D = 4 center vortices are closed flux surfaces, which 
act as wave guides for low-energetic quarks~\cite{Reinhardt:2002cm} 
and their percolation very likely triggers the condensation
of quarks. Their topological charge can be understood in terms of
their intersection number~\cite{Engelhardt:1999xw,Reinhardt:2001kf} 
or, when center vortices are 
considered as time-dependent closed loops in 3-dimensional space, by their 
writhing number~\cite{Reinhardt:2001kf}. 

For the study of long range structures and topological excitations of the QCD 
vacuum, quantum fluctuations pose a serious challenge in the analysis of
lattice configurations. These hard UV modes dominate the action and 
the long range structures are hidden under short distance noise. 
In recent years it has been 
understood that the low-lying eigenmodes of the lattice Dirac operator 
provide a natural filter sensitive to long range structures. These 
low-lying modes can be studied and information on the underlying IR 
structures of the gauge field can be extracted. 

In this article we study the connection between center vortices and 
properties of Dirac eigenmodes for quenched SU(2) configurations.
In particular we apply the technique of Ref.\ \cite{deForcrand:1999ms} 
to remove the center vortices. 

A central question of our article is how the removal of center vortices 
affects the long range topological 
structures of the Yang-Mills vacuum. As outlined,
this question can be addressed through an analysis of the 
low-lying Dirac eigenmodes. We will show in this article that 
topological modes are destroyed when removing
the center vortices. In particular the zero modes are gone completely. 
When analyzing the low-lying modes with non-vanishing eigenvalues, we 
find that their local chirality is gone and they do not resemble 
small perturbations of instanton-type zero modes. 

For the spectrum of the Dirac operator we show that in many respects 
the spectrum of vortex-removed configurations resembles the spectrum 
of free fermions. This implies that the vortex-removed configurations 
cannot build up a non-vanishing chiral condensate via the Banks-Casher 
relation. 

A legitimate criticism of removing the vortices is the fact that this 
procedure is a quite drastic modification of the gauge field. In this
article we implement a crucial test by applying random changes to the 
original gauge configurations. In particular we multiply randomly 
chosen links of the lattice with the non-trivial center element. 
We demonstrate 
that to a large extent the topological information is stable under 
such random changes. 
The number of zero modes is essentially invariant and also the 
chiral properties of zero-modes and near zero-modes. Our test shows 
that the center vortices are correlated in a highly non-trivial way
and their removal specifically destroys topological features of the 
gauge field configuration. 

Finally we will demonstrate for smooth vortex configurations that they
indeed give rise to zero modes of the Dirac operator. We analyze
the procedures of center projection and vortex removal for these 
configurations. The findings for the smooth configurations 
support our interpretation of the results for center projection 
and vortex removal applied to thermalized configurations.

Our article is organized as follows: In the next section we discuss technical
aspects of our calculation, in particular the preparation of the original,
the vortex-removed and the random-changed configurations, as well as the 
Dirac operator and the computation of the eigensystem. In Section 
3 we discuss eigenvalue spectra and their interpretation. In Section 4
we present the observables based on the Dirac eigenvectors. In 
Section 5 we study Dirac spectrum and eigenmodes for the 
hand-constructed smooth vortex configurations.
Our paper closes with a summary and the discussion of the results.


\section{Technicalities}

\subsection{Preparation of the original ensembles}

The SU(2) lattice configurations were generated with 
the standard heat bath algorithm using the Wilson action. 
After a careful thermalization, each ``measurement'' was taken 
after 20 dummy heat bath sweeps. We analyze a total of 100 quenched 
SU(2) configurations generated on a $12^4$ lattice at $\beta = 2.5$. 


\subsection{Center vortices and their removal}
\label{sect:cvortices}

The method of identification \cite{DelDebbio:1996mh} and removal 
\cite{deForcrand:1999ms} of center 
vortices is based on the so-called maximal center gauge (MCG) 
\cite{DelDebbio:1996mh,DelDebbio:1998uu}. 
If 
\begin{equation} 
U^\Omega _\mu (x) \; = \; \Omega (x) \; U_\mu (x) \; 
\Omega ^\dagger (x+\mu ) \; , \hbo 
\Omega (x) \in \mathrm{SU(2)} 
\end{equation}
denotes the gauge transformed link, MCG fixing is implemented by 
maximizing 
\begin{equation} 
S_{\mathrm{fix}} \; = \; 
\sum _{x, \mu } \Bigl[ \tr U^\Omega _\mu (x) \Bigr]^2 
\; \stackrel{\Omega }{\rightarrow } \; \mathrm{max } 
\end{equation}
with respect to $\Omega (x)$ thereby bringing each link as close as
possible to a center element \cite{DelDebbio:1998uu},
or a given gauge field configuration as close as possible to a collection 
of center vortices~\cite{Engelhardt:1999xw}. This gauge condition 
was implemented by using an iteration-overrelaxation (IO) procedure 
(details are presented in~\cite{DelDebbio:1998uu}). 
The IO procedure was stopped when the difference between 
the variational action $S_{\mathrm{fix}}$ of two subsequent gauge 
fixing sweeps was smaller than $10^{-6}$. 

After MCG fixing the center vortices are identified by replacing each
link $U^\Omega_\mu (x)$ by its closest center element 
$Z_\mu (x) \in Z(2)$, thereby each gauge configuration $U_\mu (x)$ is
converted into an ``ideal'' center vortex configuration, consisting of
closed surfaces of plaquettes being equal to a non-trivial center
element $Z_\mu \neq \Id$. The center vortex-removed theory 
\cite{deForcrand:1999ms} is defined by replacing the original gauge fixed links 
$U^\Omega_\mu (x)$ by 
\begin{equation} 
U^\Omega_\mu (x) \; \rightarrow \;  
Z^\ast_\mu (x) U^\Omega_\mu (x) \; ,
\end{equation} 
where the $Z_\mu (x)$ are the center projected counter parts  
of the $U^\Omega_\mu (x)$. Let us emphasize that the vortex removal
procedure \cite{deForcrand:1999ms} de facto removes the center projected image 
$Z_\mu (x)$ from the original gauge fixed configuration 
$U^\Omega_\mu (x)$.


\subsection{Configurations with random changes}
\label{sect:randconfigs}

One has to keep in mind that removing the vortices is a drastic alteration of
the gauge configuration: A large portion of the links is flipped, 
i.e.\ the links are multiplied by a factor of $-1$. This change leads to a 
flip of about 3\% of the plaquettes. The average plaquette 
drops by about 5 \%, from 0.651 on the original configuration down to 0.621
after removing the vortices. It must be stressed, that the 
non-trivial center elements are highly correlated beyond the fact 
that the vortices form closed surfaces on the dual lattice. An important 
question is what the effects of a similar, but unstructured alteration 
of the lattice configuration will be.

To address this question we analyze the effect of 
uncorrelated random changes. The procedure 
for these random changes is somewhat similar to the prescription for
removing the vortices. We first randomly choose a certain 
number $N_{flip}$ of links $(x,\mu)$ where the random changes are implemented.
For the selected links we change the corresponding link variables 
$U_\mu(x)$ according to 
\begin{equation}
U_\mu(x) \; \rightarrow \; - U_\mu(x) \; ,
\label{linkrandom}
\end{equation}
i.e., we flip a randomly selected set of $N_{flip}$ link variables. 
The number $N_{flip}$ of randomly chosen links subject to the 
transformation (\ref{linkrandom}) is a free parameter. 

The number of flipped plaquettes increases quickly with the number 
$N_{flip}$ of randomly flipped links. We find that randomly flipping 
$N_{flip} = 1000$ links on 
our $12^4$ lattice (this is 1.2\% of the total number of links) leads to 
a flip of about 4.6\% of the plaquettes. This decreases the average plaquette
by about 10\%, from 0.651 on the original configuration to 0.590 after 
the random changes. 

We remark that a simple combinatorial calculation gives the 
fraction $p$ of flipped plaquettes when flipping a fraction $q$ 
of links as $p = 4q - 12q^2 + {\cal O}(q^3)$. Setting $q = 0.012$ (= 1.2 \%),
the value for our random changed ensemble, one finds $p = 0.046$ (= 4.6 \%), 
as we observe. Since the links are chosen randomly, in leading order 
the flipped plaquettes are unbiased. Thus we can estimate the average
plaquette after the random changes to be $(1 - 2p) 0.651 = 0.591$, 
almost the number we observe (0.590). If on the other hand one inserts 
$p = 0.03$ (= 3 \%), the value for vortex removal, one 
expects an average plaquette of 0.612. The discrepancy of this number to
the observed value of 0.621 for vortex removed configurations, is due to 
a non-trivial correlation of the vortices.   


\subsection{The eigensystem of the Dirac operator}

For all four ensembles, original, center projected, vortex-removed 
and random-changed,
we computed eigenvalues and eigenvectors of the chirally improved (CI) 
lattice Dirac operator~\cite{Gattringer:2000js,Gattringer:2000qu}. 
The CI 
operator is a systematic approximation of a solution of the Ginsparg-Wilson 
equation which governs chiral symmetry on the lattice. The approximation is 
ultra-local and so allows for good chiral properties at a relatively low cost. 
The ultra-locality makes the CI operator also particularly suitable for the 
analysis of topological objects. In~\cite{Gattringer:2001cf} it was e.g.\
shown that the CI operator provides a better lattice image of the 
continuum zero mode of instantons small in lattice units, 
when compared to the exactly chiral, but non-ultra-local overlap operator. 

The actual calculation of the eigensystem was done with the implicitly 
restarted Arnoldi method \cite{arnoldi}. For each configuration we calculated 
the 50 smallest eigenvalues and the corresponding eigenvectors. 
We vary the boundary conditions of the Dirac operator which is a 
powerful tool to probe the 
system~\cite{Gattringer:2002wh}-\cite{Gattringer:2004va}.
While we keep periodic boundary conditions
for the spatial directions, we allow for an arbitrary phase in the temporal 
direction. Thus the eigenvectors $\vec{v}$ obey 
\begin{equation}
\vec{v}(x + L \, \widehat{i}\,) \; = \; \vec{v}(x) \; , \; i = 1,2,3 
\; \; \; ; \; \; \; 
\vec{v}(x + L \, \widehat{4}\, ) \; = \; e^{i2\pi \varphi} \, \vec{v}(x) \; ,
\label{boundarycond}
\end{equation} 
where $L$ is the size of our lattice and $\widehat{i}, \widehat{4}$
are the unit-vectors in the spatial, respectively the temporal directions. 
The phase factor $\varphi$ can assume values between 0 and 1, 
and the special cases of $\varphi = 0$ and $\varphi = 1/2$ correspond
to periodic, respectively anti-periodic temporal boundary conditions. 

For a Dirac operator which obeys the Ginsparg-Wilson equation, the 
eigenvalues are restricted to the so-called Ginsparg-Wilson circle, a 
circle with radius 1 and center $(1,0)$ in the complex plane. For the CI 
operator, which is an approximate Ginsparg-Wilson operator, the eigenvalues
fluctuate around the Ginsparg-Wilson circle. In particular also the zero
modes do not have eigenvalues that are exactly zero. In general they
have a small, but non-vanishing real part. We stress, however, 
that there is no mixing of zero modes and near zero modes, 
since eigenvectors that correspond to 
eigenvalues with non-vanishing imaginary parts have an exactly vanishing matrix
element with $\gamma_5$ and can be identified unambiguously. Thus 
eigenmodes with real eigenvalues are the would-be zero modes of the continuum. 
 

\section{Spectra of the Dirac operator}

\subsection{Spectra for center-projected configurations}

\begin{figure}[t]
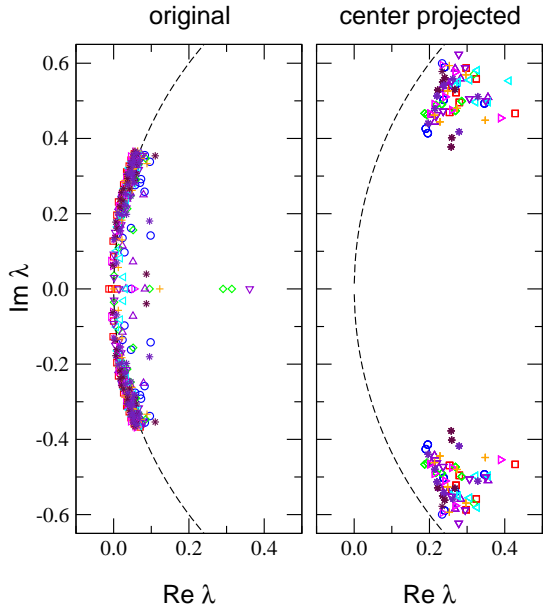

\begin{center}
\epsfig{file=spectrum_orig.eps,height=8cm,clip}
\epsfig{file=spectrum_center.eps,height=8cm,clip}
\end{center}
\vspace{-5mm}
\caption{{\sl Dirac eigenvalues $\lambda$ in the complex plane. We compare the
spectrum for the original ensemble $($lhs.\ plot$)$ 
to the spectrum for center-projected
configurations $($rhs.\ plot$)$. Each plot contains the 50 
smallest eigenvalues for
10 different configurations with every configuration represented by a 
different symbol. Anti-periodic temporal boundary conditions were used
for the Dirac operator.}
\label{spectranaive}}
\end{figure}

As a first approach to analyzing the relation between center vortices and 
Dirac eigenmodes one can calculate the 
spectrum directly for the center projected 
configurations. The result of such a calculation is shown in Fig.\ 
\ref{spectranaive}, where in the right-hand side (rhs.) plot we superimpose 
spectra for 10 center projected configurations and compare 
them to the corresponding 
spectra of the original configurations (lhs.\ plot). All spectra were 
calculated using anti-periodic boundary conditions. 

In the lhs.\ plot for the original configurations
one clearly sees that the eigenvalues cluster near the Ginsparg-Wilson circle 
(represented by the dashed curve) and only slightly scatter away from it 
(except for some real modes). 
The density of eigenvalues extends all the way to the origin indicating, that 
chiral symmetry is broken according to the Banks-Casher 
formula~\cite{Banks:1979yr}. In addition we 
find several real eigenvalues, i.e.\ would-be zero modes corresponding to 
non-vanishing topological charge via the index theorem.

The spectra for the center projected configurations on the rhs.\ plot show a
completely different picture. The spectra have developed a large gap, 
and all real
eigenvalues are gone. The eigenvalues are concentrated in two clusters 
(symmetric with respect to reflection on the real axis). These clusters
are shifted to large values of the imaginary parts, beyond the range for the
eigenvalues for the original configurations.  
This hard to interpret outcome is not really a surprise: 
The Dirac operator, which contains the gradient operator, can detect only 
smooth topologically non-trivial structures. 
The center projected configurations 
are, however, maximally discontinuous. The link variables can only 
jump from $+\Id$ to $-\Id$ when going from
a link to its neighbors. Obviously, the spectrum of the Dirac operator,
which is to a high degree determined by 
topological properties of the gauge field,  
is sensitive to the discontinuity of the link variables. 
Thus we conclude that
analyzing the Dirac spectrum directly for center projected configurations is 
a failed, although instructive attempt. We will come back to this issue
in Section 5.


\subsection{Spectra for vortex-removed configurations}
\begin{figure}[t]
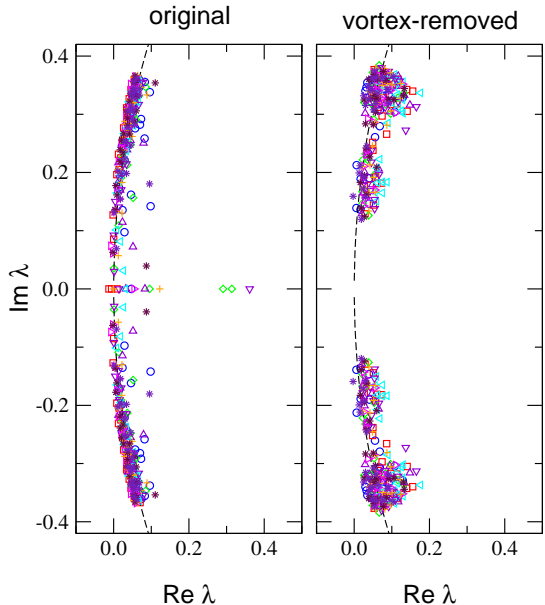

\begin{center}
\epsfig{file=spectrum_orig2.eps,height=8cm,clip}
\epsfig{file=spectrum_vrm.eps,height=8cm,clip}
\end{center}
\vspace{-5mm}
\caption{{\sl Dirac eigenvalues $\lambda$ in the complex plane. We compare the
spectrum for the original ensemble $($lhs.\ plot$)$ to 
the spectrum for vortex-removed
configurations $($rhs.\ plot$)$. Each plot contains the 50 smallest 
eigenvalues for 10 different configurations, 
with every configuration represented by a different symbol. Anti-periodic 
temporal boundary conditions were used for the Dirac operator.}
\label{spectravrm}}
\end{figure}
A way to study the role of center vortices without 
the drastic measure of projecting
the gauge links onto their center elements is the 
technique of removing the vortices
outlined in Section \ref{sect:cvortices}. 
Although this procedure also introduces 
short range disorder, the changes are by far not as drastic as for 
center projection (at least as long as the vortices are not too fat, see
below for more details), and 
one expects the approach via the analysis of Dirac eigenmodes to be viable. 

In the rhs.\ plot of Fig.\ \ref{spectravrm} we 
show superimposed the spectra for 10 
vortex-removed configurations and in the lhs.\ plot the spectra for the 
corresponding original configurations. We use the same set of configurations 
already used in Fig.\ \ref{spectranaive} (note that we have changed 
the vertical scale), and again apply anti-periodic 
temporal boundary conditions.   
Now the situation in the rhs.\ plot is quite different from the situation in 
the rhs.\ plot of Fig.\ \ref{spectranaive}. In particular the spectrum does
not display the strong shift towards large imaginary parts, and essentially 
stays in the same range as the original spectra. Also the eigenvalues are 
still quite close to the Ginsparg-Wilson circle. However, again we observe a 
gap in the spectrum, indicating that on the vortex-removed configurations the 
chiral condensate vanishes. 

A crucial observation that can be made, is the 
fact that the eigenvalues fall into clusters. It is particularly interesting  
to count the number of eigenvalues in the lower cluster (and its mirror image
obtained by reflection at the real axis). 
One finds that this occupation number 
typically is 8 eigenvalues per configuration for this lowest cluster. 

In an attempt to
understand this occupation number of the lowest cluster 
we now look at the spectrum
for the free case, which can be calculated using Fourier transform. 
Since we are only interested in the lowest eigenvalues, 
we can ignore the effects
of the periodicity of the Brillouin zone caused by the lattice and work 
with the continuum form 
$\widehat{D}(p) = i p_\mu \gamma_\mu \times {\bf 1}_2$ 
of the Dirac operator in momentum space. 
The factor ${\bf 1}_2$ is a $2 \times 2$
unit matrix coming from the trivial SU(2) field configuration.
The eigenvalues of $\widehat{D}(p)$ are given by
\begin{equation}
\lambda \; = \; \pm i \sqrt{p_1^2 + p_2^2 + p_3^2 + p_4^2} \; .
\label{freeeigenvalues}
\end{equation}
Note that each eigenvalue is 4-fold degenerate, where a 2-fold degeneracy 
follows from the block-diagonal structure of the $\gamma$-matrices, and 
another 2-fold degeneracy from the trivial SU(2) color structure. 
For the momenta
$p_\mu$ we insert the discrete momenta allowed on the lattice. These momenta
are sensitive to the boundary conditions we use. In particular we find (compare
Eq.\ (\ref{boundarycond}) for the definition of the boundary condition 
parameter $\varphi$)
\begin{equation}
p_i = \frac{2\pi}{a N} k_i \; , \; i = 1,2,3 
\; \; , \; \;
p_4 = \frac{2\pi}{a N} (k_4 + \varphi) \; \; \mbox{with} 
\; \; k_\mu = 0, \pm 1, \pm 2 \, ... \; .
\end{equation}
Here $N$ is the total number of lattice points in 
one direction and $a$ denotes 
the lattice spacing, i.e.\ the physical extension $L$ 
of our lattice is $L = aN$. 

Let us now analyze what degeneracy of the smallest eigenvalue we find for the 
anti-periodic boundary conditions used in Fig.\ \ref{spectravrm}. For this 
case we have $\varphi = 1/2$. We find a 4-component of the momentum of 
$p_4 = \pi/aN$ for $k_4 = 0$ and $p_4 = -\pi/aN$ for $k_4 = -1$. These two 
values differ only by a sign, and since the eigenvalues (\ref{freeeigenvalues})
depend only on the square $p_4^2$ we find an extra degeneracy
of the lowest eigenvalue (defined by $p_1=p_2=p_3=0, \, p_4 = \pm \pi/aN$). 
Thus for anti-periodic boundary conditions we obtain an 8-fold degeneracy
as observed in the lowest clusters in the rhs.\ plot of Fig.~\ref{spectravrm}.

It is tempting to interpret the eigenvalues in the rhs.\ plot of Fig.\
\ref{spectravrm} as a slightly disturbed free spectrum. If indeed removing the
vortices removes all topological excitations, then one would expect that the 
remaining configuration is essentially a trivial configuration plus some 
fluctuations. In order to test the hypothesis, that after removing the vortices
one is essentially left with a free spectrum, we now use different
values for the boundary condition parameter $\varphi$. 
For a value of $\varphi = 0$ (periodic boundary conditions) we again obtain an 
8-fold degeneracy, since for $p_1=p_2=p_3=p_4 = 0$ the two signs in 
(\ref{freeeigenvalues}) both give $\lambda = 0$. 
For the case of $\varphi = 1/4$ 
we obtain only 4-fold degeneracy of the lowest eigenvalue characterized by 
$p_1 = p_2 = p_3 = 0, \, p_4 = \pi/(2 a N)$. 
  
\begin{figure}[t]
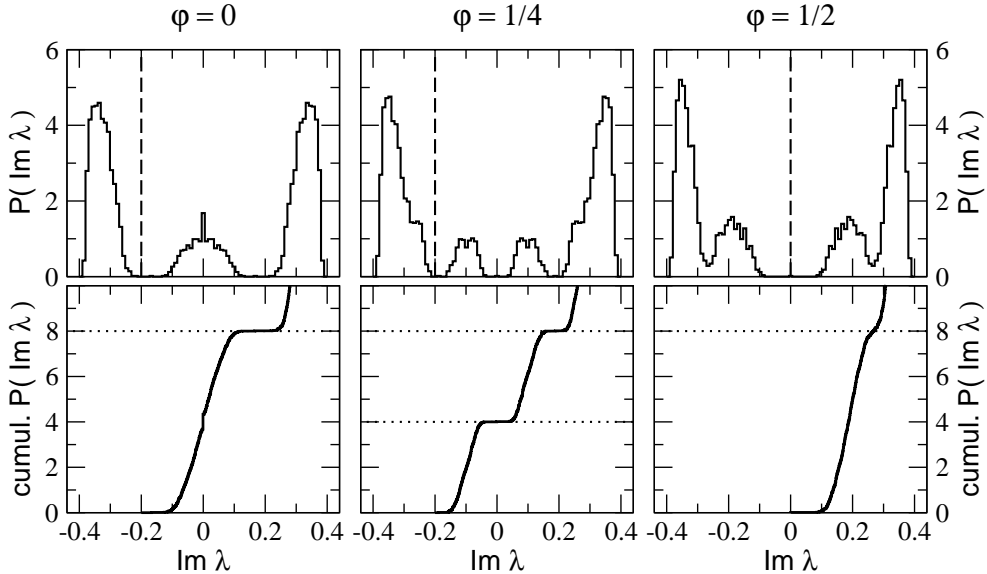

\begin{center}
\epsfig{file=histo_vrm000.eps,height=7.6cm,clip}
\hspace{-2mm}
\epsfig{file=histo_vrm025.eps,height=7.6cm,clip}
\hspace{-2mm}
\epsfig{file=histo_vrm050.eps,height=7.6cm,clip}
\end{center}
\vspace{-5mm}
\caption{{\sl Histograms $($top row$)$ and cumulated histograms 
$($bottom row$)$ 
of the imaginary parts of Dirac eigenvalues for vortex-removed configurations. 
We compare the histograms for periodic $(\varphi = 0)$ 
b.c.\ $($lhs.\ plots$)$, for $\varphi = 1/4$ $($center plots$)$
and for anti-periodic $(\varphi = 1/2)$ b.c.
The vertical dashed lines in the top row plots show where we started with
summing the numbers for the cumulative histograms. In the cumulative 
histograms 
$($bottom row$)$ we use dotted horizontal lines to indicate 
where the cumulative histogram exhibits a shoulder.}
\label{histovrm}}
\end{figure}
To make our study more quantitative we present our results for the 
spectra now as histograms using all 100 configurations which we analyzed. 
In particular we use two types of histograms: Histograms where the density of 
the eigenvalues is plotted as a function of Im$\lambda$, and cumulative 
histograms where the total number of eigenvalues below a certain value of 
Im$\lambda$ 
is shown. For the vortex-removed configurations the results are shown in Fig.\
\ref{histovrm} with the regular histograms in the top row and the 
corresponding 
cumulative histograms in the bottom row. The three columns of plots are for
the three different boundary conditions we used (from left to right:
$\varphi = 0, \varphi = 1/4, \varphi = 1/2$).

The three regular histograms in the top row are symmetric with respect to 
reflection at the origin due to the corresponding symmetry of the spectrum. 
The histograms show a clear separation of several maxima, with the position 
of the maxima changing as a function of $\varphi$. The position of the 
lowest maxima is qualitatively at the same position 
as for the free case where  
$\lambda = 0$ for $\varphi = 0$,
$\lambda = \pm i \pi/(2aN)$ for $\varphi = 1/4$ and $\lambda = \pm i \pi/aN$
for $\varphi = 1/2$. We remark that the 50 eigenvalues, available for
each of our configurations, are not enough to obtain all eigenvalues
in the cluster corresponding to the second-smallest free eigenvalue. For
example in the case of $\varphi = 0$ the second-smallest eigenvalue,
characterized by one of the $p_\mu$ being $2 \pi/aN$, the others being 
equal to 0, is already 32-fold degenerate. Since this eigenvalue comes as a 
complex conjugate pair,
we find a total of $8 + 2 \times 32 = 72$ eigenvalues in the two smallest
eigenvalues (the larger one being a complex conjugate pair). Thus with 
our 50 eigenvalues the second cluster is not completely filled and the larger 
peaks in the top row of histograms in Fig.\ \ref{histovrm} correspond to only
a subset of eigenvalues expected. 

Let us now analyze the occupation numbers in the lowest 
cluster for the different
boundary conditions. We start summing our cumulative histograms in the minimum 
below the clusters we are interested in. In the top row plots of
Fig.\ \ref{histovrm} we mark this position by a vertical dashed line. 
The corresponding cumulative histograms are shown right below the regular 
histograms. For the case of $\varphi = 0$ we find that the cumulative histogram
shows a pronounced shoulder at a value of 8 (the cumulative histogram was 
normalized by the total number of configurations). This shows clearly that 
the lowest cluster of eigenvalues has an occupation number of 8,
matching the degeneracy of the lowest eigenvalue in the free case. 
For $\varphi = 1/4$ the histogram shows shoulders at 4 and 8, again matching 
the degeneracies of the free case. Finally for $\varphi = 1/2$ we do not see a 
clear shoulder but at least a pronounced dip in the curve. The reason is that 
for this case the lowest and the second eigenvalue are relatively close to 
each other and we do not have a clean separation of the two clusters, as 
can already be seen in the corresponding regular histogram. 

To summarize, the relative 
position of the peaks in the regular histograms, as well as the occupation 
number in each cluster as obtained from the cumulative histograms, leads 
to the conclusion that the spectrum for vortex-removed configurations 
strongly resembles a free spectrum, slightly perturbed by fluctuations. 
This finding has two important physical implications:
Firstly, we conclude that no topological objects are left in the gauge 
configurations after center vortices have been removed. The few zero modes 
(i.e., real modes on the lattice) that are found 
for $\varphi = 0$ (see Fig.~\ref{histovrm}) are the trivial zero 
modes present also for the free Dirac
operator. For the other boundary conditions the complete absence of zero 
modes already shows the absence of topological charge via the index theorem. 
Secondly, the fact that the spectral density near the origin behaves
like in the free case shows that the chiral condensate vanishes for 
vortex-removed configurations which strongly supports the findings of 
\cite{deForcrand:1999ms,Alexandrou:1999vx}.

\begin{figure}[t]
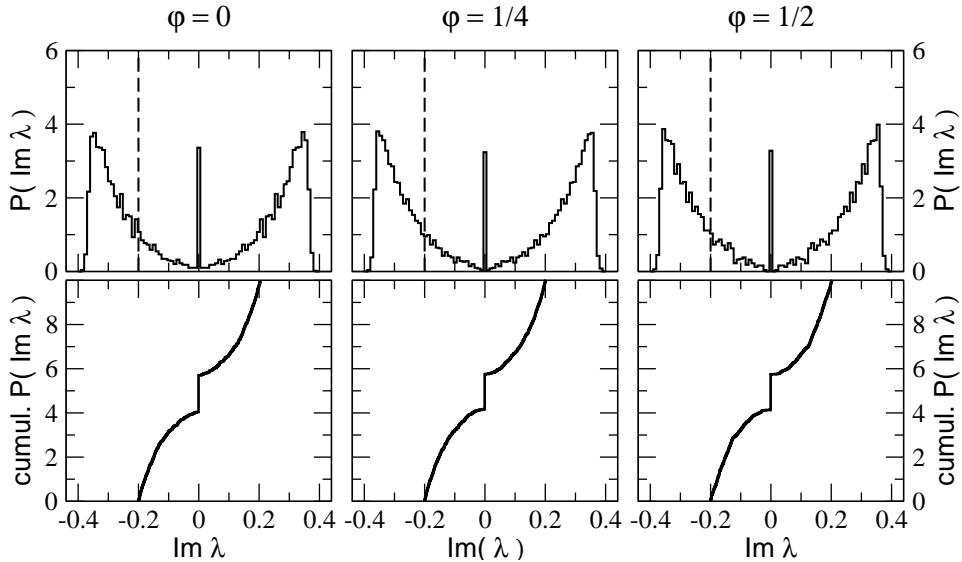

\begin{center}
\epsfig{file=histo_raw000.eps,height=7.4cm,clip}
\hspace{-2mm}
\epsfig{file=histo_raw025.eps,height=7.4cm,clip}
\hspace{-2mm}
\epsfig{file=histo_raw050.eps,height=7.4cm,clip}
\end{center}
\vspace{-5mm}
\caption{{\sl Same as Fig.\ \protect{\ref{histovrm}} but now for the 
original configurations.}
\label{historaw}}
\end{figure}
For completeness we show the histograms and cumulative histograms also 
for the original configurations (Fig.\ \ref{historaw}). The regular 
histograms in the top row do not show any pronounced peaks. The single tall 
bin at the origin is due to the zero modes which are present in the original
configurations. The histograms do not show any strong dependence on the 
boundary condition. The summation for the cumulative histograms was started 
at Im$\lambda = -0.2$ for all boundary conditions. Furthermore,
the cumulative histograms are independent of $\varphi$ and the 
only feature is the vertical step at Im$\lambda = 0$ due to the 
contribution of the zero modes. 


\subsection{Spectra for random-changed configurations}

In Section \ref{sect:randconfigs} we have discussed the random-changed 
configurations which we prepared to analyze the effect of an uncorrelated 
change of the gauge configuration. For $N_{flip} = 1000$ flipped links
we find that the action goes up twice as much as when removing the vortices 
and it is interesting to see what the effects on the spectrum turn out to be. 

\begin{figure}[t]
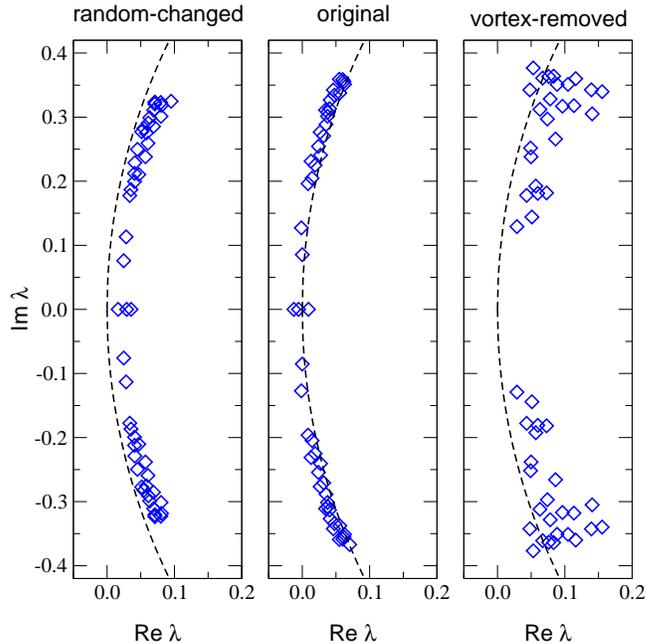

\begin{center}
\epsfig{file=singspect_rand.eps,height=8.5cm,clip}
\epsfig{file=singspect_orig.eps,height=8.5cm,clip}
\epsfig{file=singspect_vrm.eps,height=8.5cm,clip}
\end{center}
\vspace{-5mm}
\caption{{\sl Dirac eigenvalues $\lambda$ in the complex plane. We compare the
spectrum for the random-changed configuration $($lhs.\ plot$)$, 
the original configuration $($center plot$)$ and the 
vortex-removed configuration $($rhs.\ plot$)$. The random changes, as well 
as the procedure of vortex-removal, both started from the same
original configuration - the one we also use in the center plot. Anti-periodic 
temporal boundary conditions were implemented.}
\label{spectrarand}}
\end{figure}
In Fig.\ \ref{spectrarand} we compare the spectra for a random-changed 
configuration (lhs.\ plot) to the spectrum for the original configuration
(center plot) and the corresponding vortex-removed configuration (rhs.\ plot). 
We used anti-periodic temporal boundary conditions for all three cases.
When comparing the spectra of the random-changed 
and the original configurations, 
one finds that the former is slightly shifted away 
from the Ginsparg-Wilson circle
and slightly compressed in the vertical direction. 
Much more important, however, is the fact that the gross
features of the two spectra are nearly identical. In particular the number of
real eigenvalues, i.e.\ the number of would be zero modes is invariant, 
and also
the pattern of the relative spacing of the eigenvalues is very similar. This
finding does not only hold for a single configuration, 
but also bulk observables 
such as the histograms and cumulative histograms for 
random-changed configurations
are very similar to their original counterparts shown in Fig.\ \ref{historaw}. 

\begin{figure}[t]
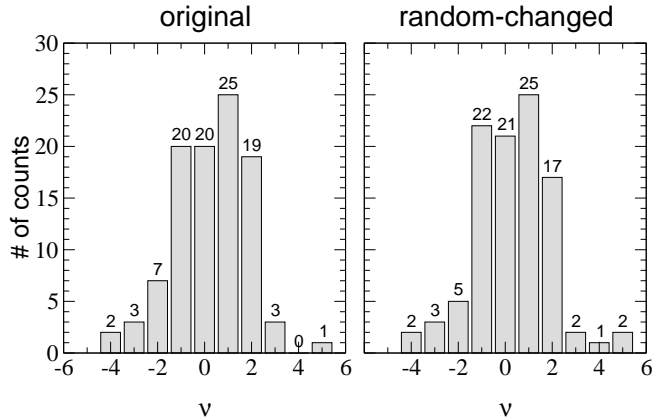

\begin{center}
\epsfig{file=qhisto_orig.eps,height=5.5cm,clip}
\epsfig{file=qhisto_rand.eps,height=5.5cm,clip}
\end{center}
\vspace{-5mm}
\caption{{\sl Distribution of the topological charge $\nu$. 
We compare data from the 
original ensemble $($lhs.\ plot$)$ to the random-changed
ensemble $($rhs.\ plot$)$.} 
\label{qhisto}}
\end{figure}
We also analyzed the distribution of the topological charge
$\nu$ in the original and the random-changed ensemble. The topological charge
$\nu$ was determined from the index theorem using the 
difference of the numbers of
left-handed and right handed zero modes. The results are displayed in Fig.\ 
\ref{qhisto} where we show the distribution of $\nu$ for the original 
configurations in the lhs.\ plot and for the random-changed configurations
in the rhs.\ plot. Again we observe that the changes 
due the random alterations of
the gauge configuration are minimal. This confirms an earlier 
finding~\cite{Gattringer:1997ci}
where it was demonstrated that the topological features
of the spectrum of the Wilson Dirac operator are quite stable when adding 
random noise. 

There are certainly many different ways of altering the original 
configuration for comparison of the effects to a removal of center vortices. 
Such alternative prescriptions could e.g.\ be inspired by a particular picture 
of QCD vacuum excitations. We remark that currently we are exploring such 
alternative modifications of the gauge field and their effect on the Dirac 
eigensystem. The random changes we use here are merely intended
to demonstrate that a non-correlated change, which increases the action 
considerably more than removing the vortices, leaves the topological features 
invariant.


\section{Results for the Dirac eigenmodes}

\subsection{Scalar density for zero and near zero modes}

After having explored the spectrum for center projected, 
vortex-removed and random-changed
configurations, it is also interesting to study the eigenvectors of the 
Dirac operator for these ensembles. A suitable observable is the scalar 
density $\rho(x)$ of an eigenvector $\vec{v}$. It is defined by 
summing the color and Dirac indices of $\vec{v}$ at each lattice point
$x$, 
\begin{equation}
\label{eq_005}
\rho (x) \; = \; \sum_{c,d} | \vec{v} (x)_{cd} |^2 \; .
\end{equation}
Since the Dirac operator transforms covariantly under a gauge transformation,
the scalar density $\rho$ is gauge invariant. 

Since the low-lying Dirac modes
do not see large fluctuations of a gauge field, the scalar density of the
low-lying Dirac modes can be used as a detector of (smooth) center
vortex flux, which does not rely on topological properties of the gauge
fields like their topological charge.

\begin{figure}[t]
\begin{center}
\epsfig{file=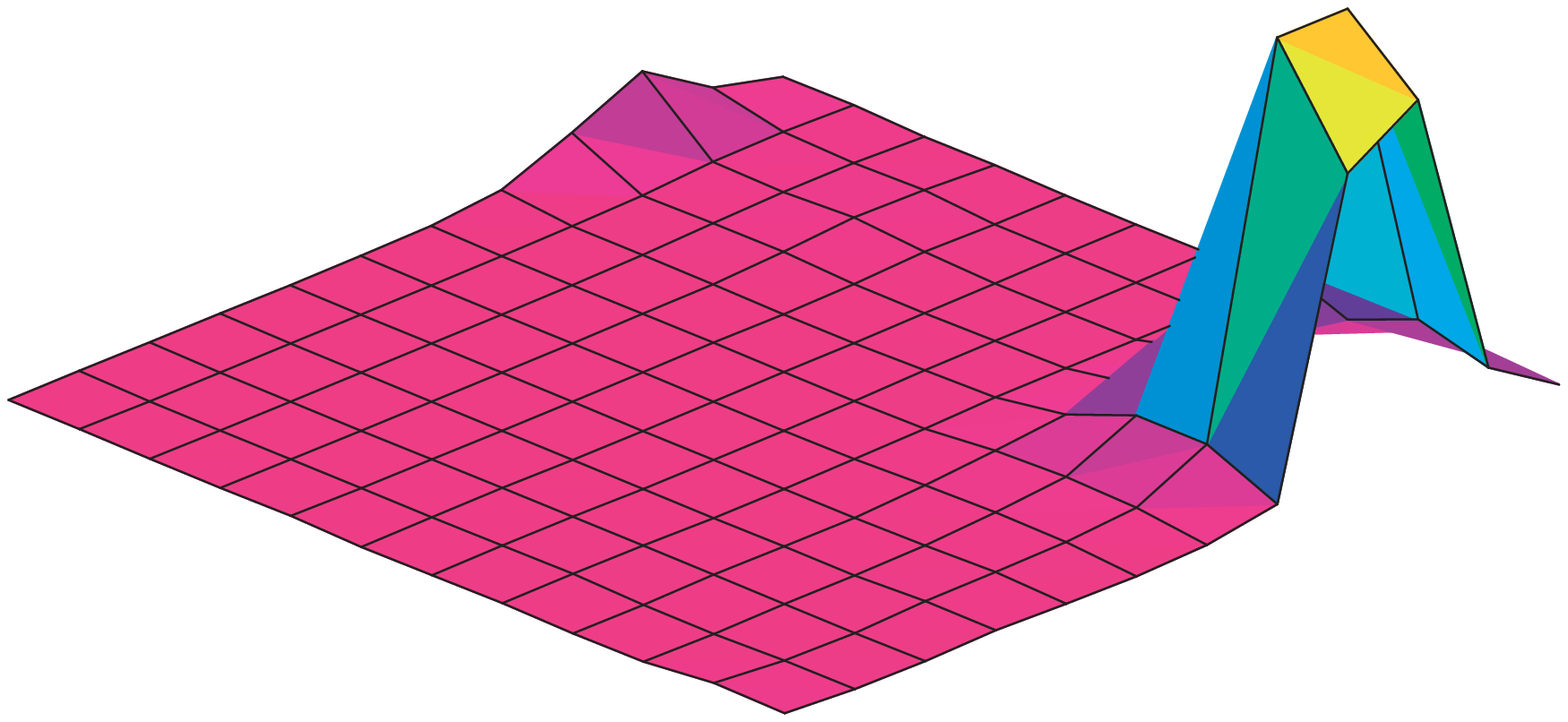,height=3cm,clip}
\epsfig{file=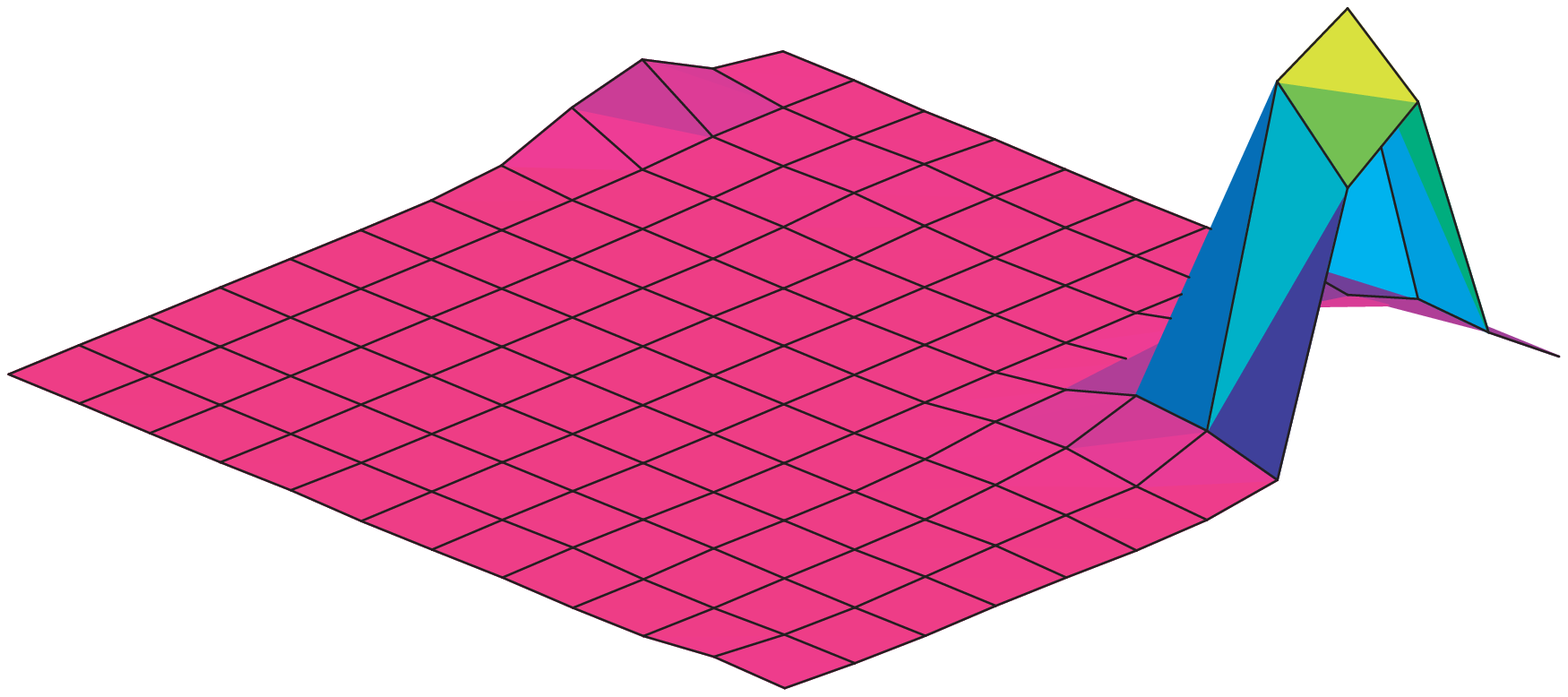,height=3cm,clip}
\end{center}
\vspace{-5mm}
\caption{{\sl Plots of the scalar density $\rho(x)$ 
for the zero-mode of the original configuration $($lhs.\ plot$)$ and the 
corresponding random-changed configuration $($rhs.\ plot$)$. 
We show the same 2-d slices for both configurations.}
\label{slicezero}}
\end{figure}
In Fig.~\ref{slicezero} we show the scalar density over a 2-d slice 
through the lattice for a zero mode. The lhs.\ plot is for the original 
configuration, the rhs.\ plot after random changes. The slice was 
chosen such, that it cuts through the maximum of $\rho(x)$. We 
find a strongly localized lump, which, in an orthodox interpretation, 
would be the zero mode due to an instanton in the underlying 
gauge field\footnote{ Later we will identify lumps of fractional topological 
charge with vortex intersection points.} (see Refs.\ \cite{Horvath:2003yj,
Horvath:2002yn, Aubin:2004mp} for 
an alternative picture.) 
 The lump is essentially unchanged by 
random changes. Since for the vortex-removed configurations all 
the zero modes are gone, we cannot produce an equivalent plot 
for this ensemble. 

However, it has long been known that also the ``near zero modes'', 
corresponding
to eigenvalues with small but non-vanishing imaginary parts, show lumpy
structures\footnote{According to the instanton picture these lumps 
originate from
instantons and anti-instantons perturbing each other. This perturbation
is, however, only weak, such that locally the near zero mode still resembles
the case of the unperturbed instanton.}. 
Since also the vortex-removed configurations do have such near zero 
modes it is possible to compare the scalar density for all four ensembles, 
original, 
vortex-removed, random-changed and center projected.

\begin{figure}[t]
\begin{center}
\epsfig{file=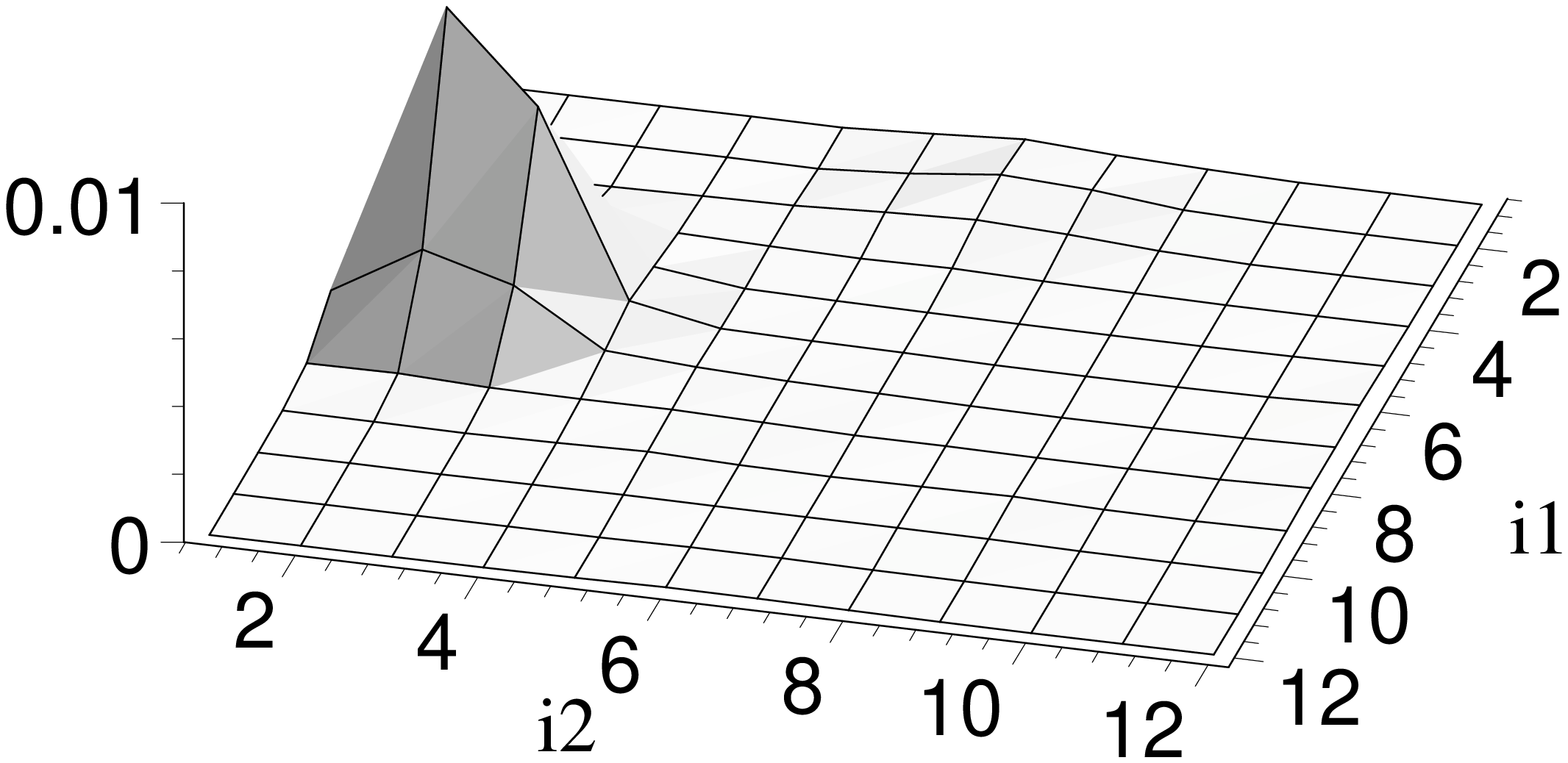,height=6cm,clip,angle=270}
\epsfig{file=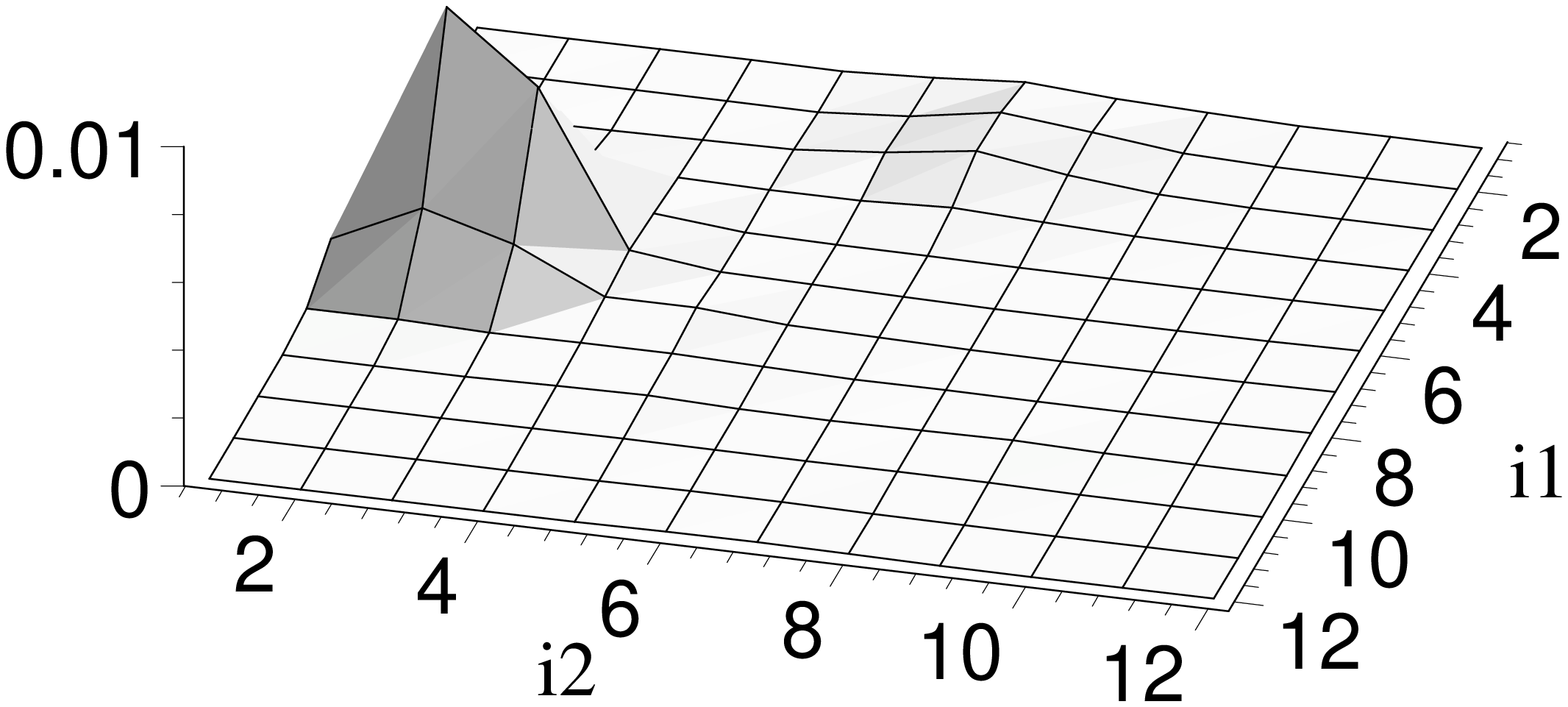,height=6cm,clip,angle=270} \\
\epsfig{file=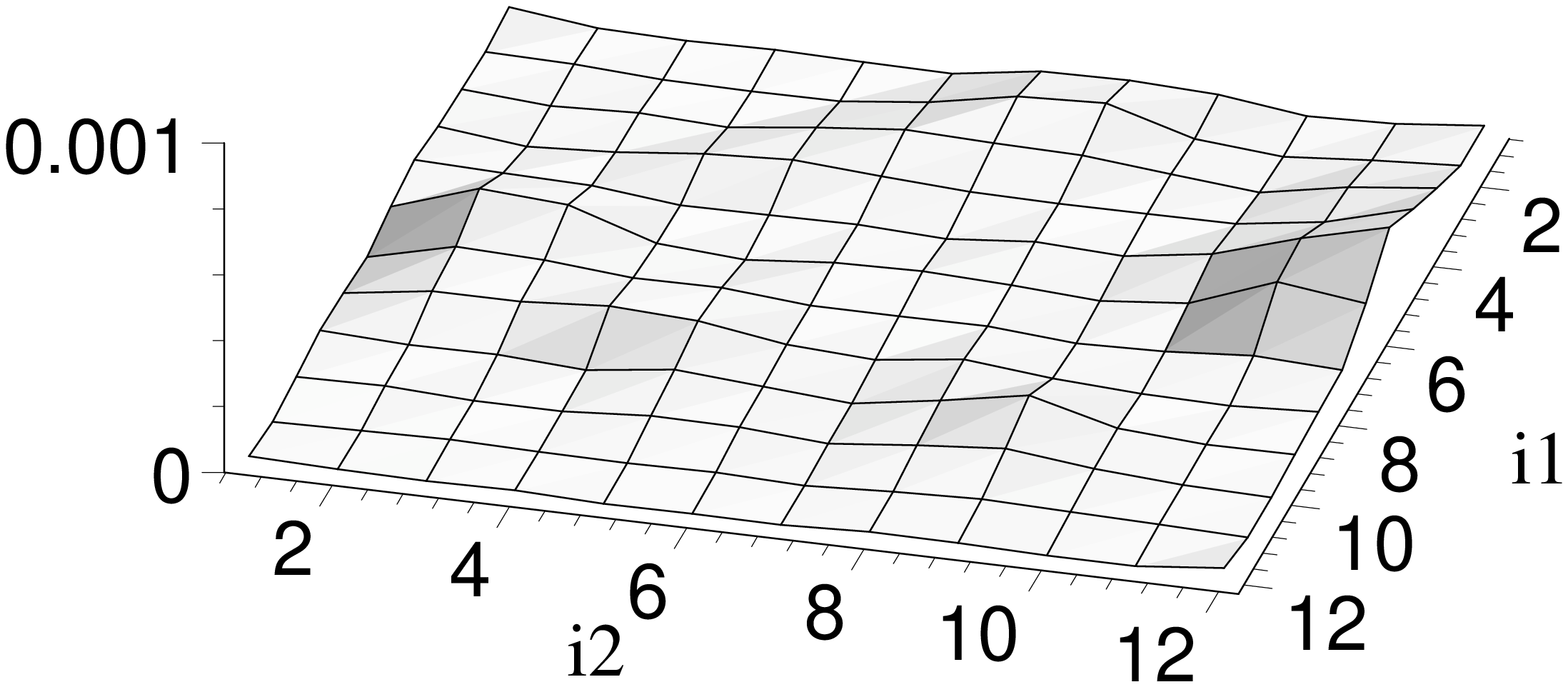,height=6cm,clip,angle=270}
\epsfig{file=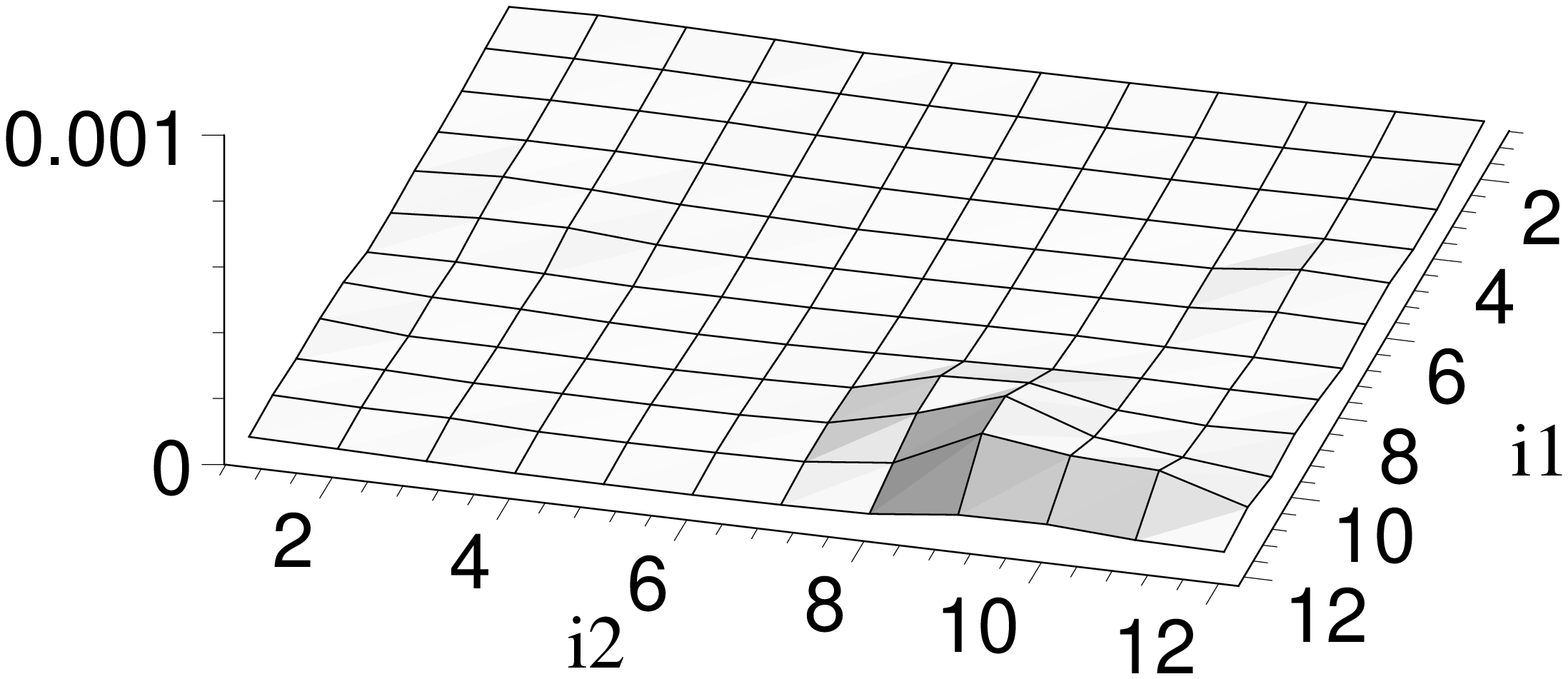,height=6cm,clip,angle=270}
\end{center}
\vspace{-5mm}
\caption{{\sl Plots of the scalar density $\rho(x)$ 
for a near-zero-mode of the original configuration $($top left$)$, 
as well as the corresponding random-changed $($top right$)$, 
vortex-removed $($bottom left$)$ and center projected $($bottom right$)$ 
configurations. We show the same 2-d slice for all four configurations.  
The vertical axis for the vortex-removed and center projected plot
$($bottom$)$ is stretched by a factor of $10$.}
\label{slicenearz}}
\end{figure}

Fig.~\ref{slicenearz} is such a comparison, with the top left plot
showing the original confi\-gu\-ration, the top right plot the 
random-changed configuration, the bottom left plot is for the vortex-removed 
and the bottom right for the center projected configuration. 
Again we show a slice through the maximum of $\rho$. The original and 
random-changed 
configurations show pronounced lumps which are located at the same
position and have essentially the same shape. When slicing the scalar density 
for the vortex-removed configuration at the same position we find that the
lump is gone completely. In the plot we even stretched the vertical scale by 
a factor of 10 to make the remaining small wiggles visible at all. Also
at other positions on the lattice we do not find localized structures in the
near zero modes of vortex-removed configurations. This holds not only
for the particular configuration used in Fig.\ \ref{slicenearz}, but is a
generic feature of the whole ensemble. Thus we must conclude, that removing 
the vortices also removes the lumpy structures in the eigenmodes. This
observation is in agreement with our findings concerning the absence of 
the chiral condensate: The density of near zero eigenvalues which, 
according to the Banks-Casher formula, is necessary
to build up the chiral condensate, comes from topological objects
which perturb each other only slightly. 
If the chiral condensate is gone for vortex-removed configurations, one 
expects also a dramatic alteration of the near zero modes. This is exactly
what we observe. 

Surprisingly, in the center projected configuration the localized lump 
is also gone. From this one might be tempted
to conclude that the topological lump seen in the original lattice
configuration is not related to center vortices at all. However, in Section 5
we will provide arguments that, very likely, the toplogical lumps are not
absent in the center projected lattice configuration but are just not
seen by the scalar density of the lowest non-zero mode. The reason is
that the center projected lattice configurations fluctuate to wildly to
be seen by a low energy filter. In Section 5 we will provide arguments 
supporting this interpretation.


\subsection{Local chirality of near zero modes}

As addressed above, the low-lying 
near zero modes are expected to locally resemble
true zero modes. In particular, they are expected to be locally chiral. 
In order to test for local chirality, one can introduce chiral (right- and
left-handed) densities $\rho_+(x), \rho_-(x)$, which are 
obtained by projecting the scalar density $\rho(x)$ to a specific chirality,
\begin{equation}
\label{eq_007}
\rho_\pm(x) \; = \; \sum_{c,d,d^\prime} \vec{v}(x)_{cd}^* \, 
\frac{1}{2} [ 1 \pm \gamma_5 ]_{dd^\prime} \, \vec{v}(x)_{cd^\prime} \; , 
\end{equation}
where the subscripts denote color and Dirac indices. 
If the near zero modes are locally chiral, then for a given space-time 
point $x$ only one of the two densities $\rho_+(x), \rho_-(x)$ is non-zero. 

An observable which further analyzes 
the properties of the near zero modes is the local chirality 
observable $X$, introduced in~\cite{Horvath:2001ir,Horvath:2002gk}
and studied by several groups~\cite{DeGrand:2001pj}-\cite{Hasenfratz:2001qp}.
The local chirality variable $X$ is obtained by mapping the ratio 
\begin{equation}
r(x) \; = \; \frac{\rho_+(x)}{\rho_-(x)} \; ,
\end{equation}
which takes values in the interval $[0,\infty)$, to  
the interval $[-1,1]$,
\begin{equation}
X(x) \; = \; \frac{4}{\pi} \; 
\tan^{-1} \left( \sqrt{r(x)} \right) \; - 
\; 1 \; .
\end{equation}
If the near zero modes are locally chiral, one expects that the
distribution of $X$ shows a double peak structure with maxima near $\pm1$.
If on the other hand the near zero modes are not locally chiral, 
one expects a single peak 
near the origin. We determine this distribution using histograms
for the values of $X(x)$, where we include all those lattice points $x$ 
where we find the largest 12.5\% of the scalar density. Thus we analyze 
the local density only for the highest peaks in $\rho(x)$. 

\begin{figure}[t]
\begin{center}
\epsfig{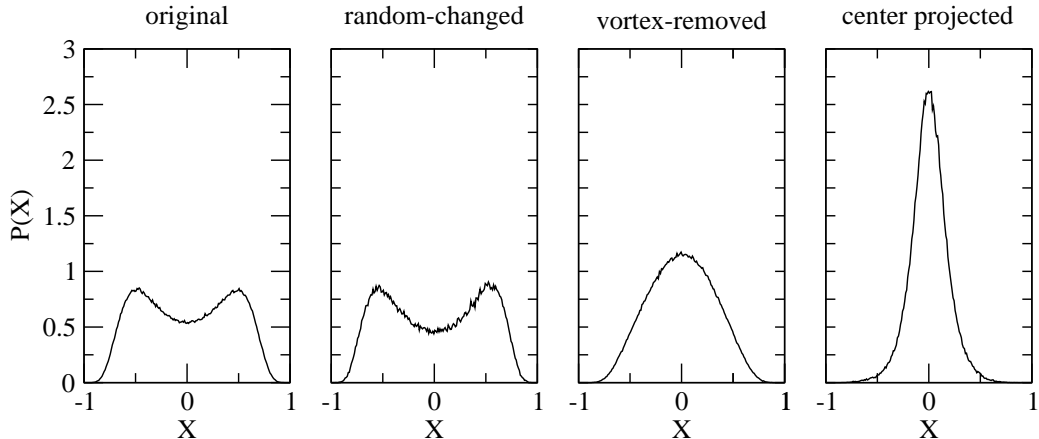}
\end{center}
\vspace{-5mm}
\caption{{\sl Distribution of the local chirality variable $X$.
We compare data $($from left to right$)$ from the 
original ensemble to data for the random-changed
ensemble, the vortex-removed and center projected configurations.}
\label{xhisto}}
\end{figure}
The results for the local chirality are shown in Fig.\ \ref{xhisto}. For the 
original and the random-changed configurations,  we 
find a clear double-peak structure indicating that the near zero modes are 
locally chiral for these two cases. For the vortex-removed configurations  
the double-peak structure is gone, indicating that the wiggles in $\rho(x)$
that remain after removing the vortices are not locally chiral structures. 
This confirms our previous interpretation of these wiggles as structureless
fluctuations. As before, we find that the 
center projected configurations are too 
singular for an analysis with the low-lying Dirac eigenmodes. 
 

\section{Results for smooth center vortices}

During the course of our analysis we have found that center 
projected configurations do not give sensible results when analyzed with 
the low-lying Dirac eigenmodes. We have argued that center projection yields 
configurations that are too singular for the eigenmode analysis. 
This is a consequence of the center projection, which converts 
the originally fat 
center vortices (present in the full Yang-Mills ensemble) into
``ideal'' center vortices, whose transversal extension is one lattice
spacing only.

However, in this Section we demonstrate that sufficently smooth center 
vortices do indeed produce zero modes and thus 
can also give rise to chiral symmetry breaking. For this study we 
use smooth vortex configurations consisting of two pairs of parallel 
smeared out planar 
vortex sheets, intersecting perpendicularly in four (smeared out) 
intersection points, 
each carrying the topological charge $\pm \frac{1}{2}$ (for more details 
see Ref.~\cite{Reinhardt:2002cm}). 
The vortex sheets are closed by the periodic boundary 
condition on the 4-torus. The orientation of the vortex sheets, i.e.\ the 
direction of the flux is such that the contributions from the intersection 
points add up to a total topological charge of $\nu = 2$ for configuration 
$\# 0$ and $\nu = 0$ for configurations $\# 1$, $\# 2$, $\# 3$, $\# 4$, 
respectively.


\subsection{Spectra of smooth center vortex configurations 
vs.\ center projected vortices}

\begin{figure}[t]
\begin{center}
\epsfig{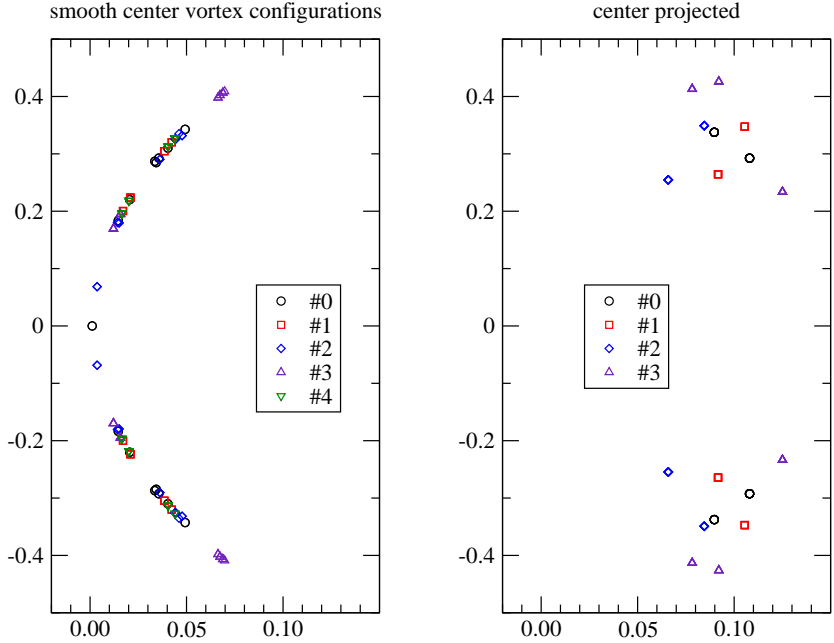}
\end{center}
\vspace{-5mm}
\caption{
\label{slice-neu-1}
{\sl Dirac spectrum for smooth center vortices
$($lhs.$)$ and their center projected counter parts $($rhs.$)$.
Note that the center projection is the same for configurations 
$\# 0$ and $\# 4$.}}
\end{figure}

Let us begin with analyzing the Dirac spectra for our smooth 
center vortex configurations. In the lhs.\ plot of Fig.~\ref{slice-neu-1} 
we show the Dirac spectra for our 5 configurations and compare them
to the spectra obtained for the corresponding center projected 
configuratons. 

The Dirac spectra of the smeared out center vortex configuration have no gap 
around zero virtuality. In particular, for configuration $\# 0$ with
non-zero topological charge $\nu = 2$ there is a zero mode, 
which is two-fold degenerate. After center projection the gap in the Dirac 
spectra emerges and the zero mode is gone. This is not surprising since in the 
process of center projection  the center vortices lose their orientation and 
thus their total topological charge, although they still carry the individual 
local spots of fractional topological charge. But these local contributions 
will usually add up to zero total charge. This is because orientable closed 
surfaces in $D = 4$ have zero total self-intersection number, which up to a
factor of $\frac{1}{4}$ represents the topological charge of center vortex
surfaces~\cite{Engelhardt:1999xw,Reinhardt:2001kf}. 

\begin{table}[t]
    \begin{center}
        \begin{tabular}{c|c|c|c|c|c|c}
          conf. & vortex 1 & vortex 2 & vortex 3 & vortex 4 & $\nu$ \\
          &$(k_0,k_3,f)$&$(k_0,k_3,f)$&$(k_1,k_2,f)$&$(k_1,k_2,f)$& 
          \\ \hline
          $\# 0$ & $(10,14,0.5)$ & $(4,14,0.5)$ & $(4,4,0.5)$ & 
          $(12,14,0.5)$ & 2 \\
          $\# 4$ & $(10,14,-0.5)$ & $(4,14,0.5)$ & $(4,4,0.5)$ & 
          $(12,14,-0.5)$ & 0 \\   
        \end{tabular} 
   \caption{\label{tab_001} {\sl Smooth center vortex
   configurations consisting of 4 planar center vortex sheets 
   which are all parallel to coordinate planes $k_i-k_j$. 
   The table shows the flux $f$ and the coordinates $(k_0,k_3)$ of the 
   two vortices parallel to the $k_1-k_2$ plane and the coordinates 
   $(k_1,k_2)$ for the two vortices parallel to the $k_0-k_3$ plane for
   the configurations $\# 0$ and $\# 4$. Configuration $\# 4$ arises
   from configuration $\# 0$ by changing the orientation of the flux.
   Furthermore, the total topological charge $\nu$ of the configuration
   is given.}}
   \end{center}

\end{table}

\begin{figure}[t]
\begin{center}
\epsfig{file=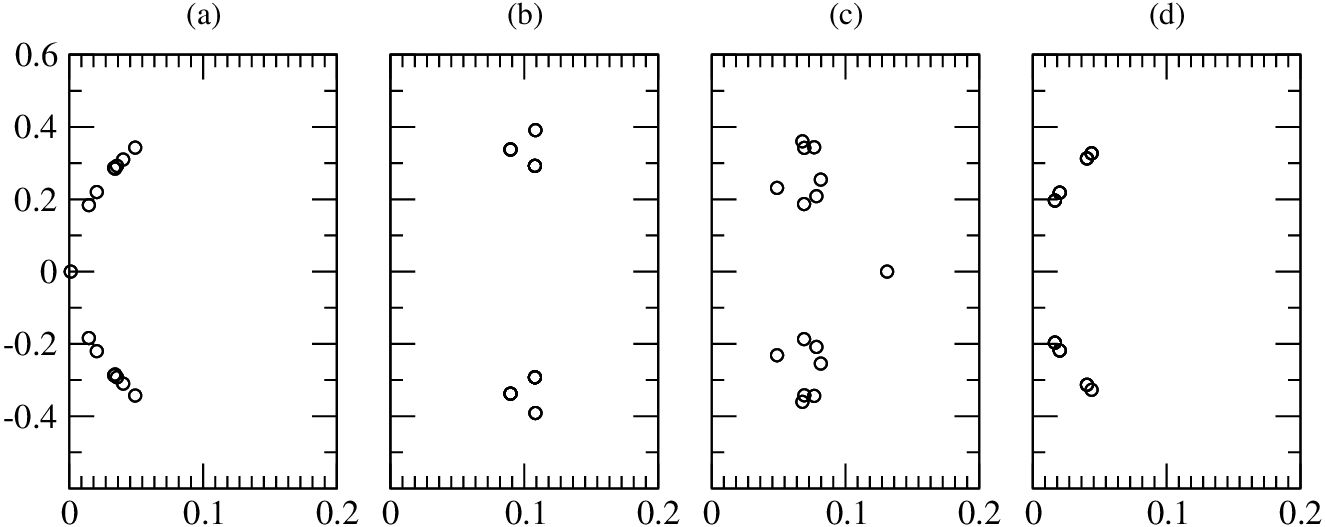,height=5.2cm} 
\end{center}
\vspace{-5mm}
\caption{\label{fig_000}{\sl Dirac spectra of $(a)$ the fat center 
vortex configuration $\# 0$ 
$(\nu = 2)$, its $(b)$ center projected and $(c)$ vortex-removed
counter part, and $(d)$ the fat center vortex configuration with $\nu=0$
obtained from $(a)$ by changing the orientation of the flux.}}
%
\begin{center}
\epsfig{file=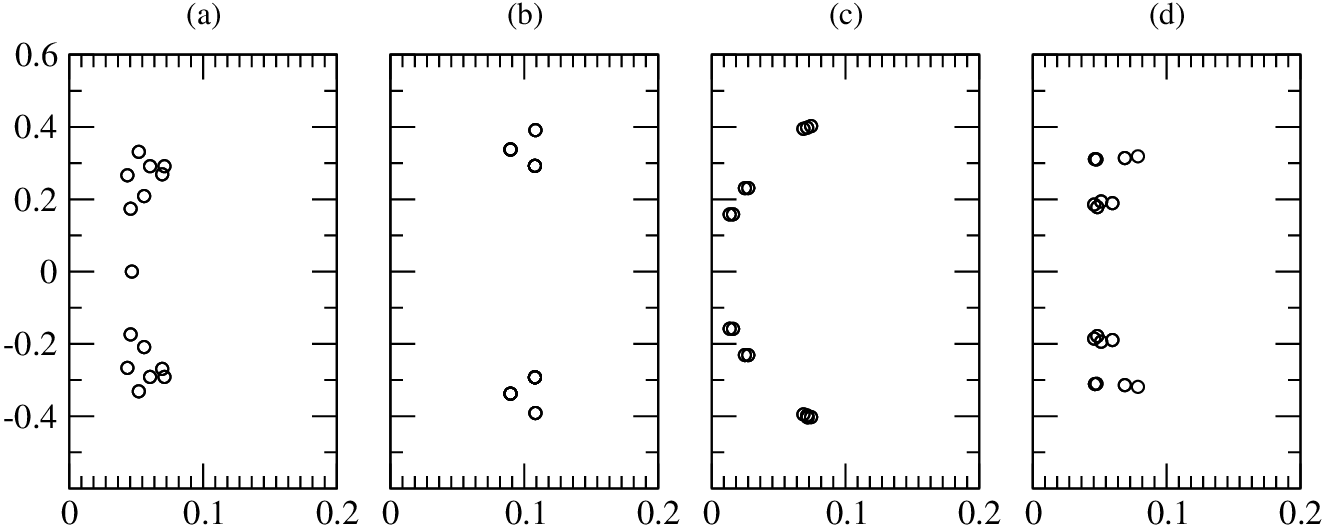,height=5.2cm} 
\end{center}
\vspace{-5mm}
\caption{\label{fig_001}{\sl Dirac spectra for the same fat vortex
configuration as in Fig.~\ref{fig_000} except that the transversal
extension of the vortex flux was shrunk to 2 lattice spacings.}}
\end{figure}

To illustrate the effect of center projection and center vortex removal
in more detail we compare in Fig.~\ref{fig_000} the Dirac spectrum of 
the fat center vortex configuration $\# 0$ (see Table \ref{tab_001}) 
with topological charge $\nu=2$ with those of its (b) center projected 
and (c) center vortex-removed counter parts. 
For sake of comparison, in Fig.~\ref{fig_000}(d) we also show the Dirac
spectrum of the fat center vortex configuration $\# 4$, which has the
same vortex surfaces as the configuration $\# 0$ (a), however, with 
different orientations of their fluxes such that $\nu = 0$ (cf.~Table
\ref{tab_001}). 
Both fat center vortex configurations $\# 0$ and $\# 4$ have the same
center projected and center vortex-removed counter parts and thus
the same Dirac spectrum shown in
Fig.~\ref{fig_000}(b) and (c). As is seen, center
projection (b) changes the Dirac spectrum as drastically as center
vortex removal (c). This is somewhat counter intuitive since the
configurations under consideration are plain vortex configurations 
but, in fact, can be easily understood: The vortex configuration
considered in Fig.~\ref{fig_000}(a), (d) are extremely fat. In the
transversal directions they stretch out over the whole lattice.
Accordingly, there are only small gradients involved and our lattice
Dirac operator works well: all eigenvalues are on the Ginsparg-Wilson
circle, see fig \ref{fig_000}(a), (d). This property is lost in the
center projection, which replaces the originally fat vortices by the
ideal (very thin) ones, whose transversal extension is one lattice
spacing. Hence the center projected configuration contains large
gradients which cannot be properly captured by our Dirac operator.
Accordingly its eigenvalues are off the Ginsparg-Wilson circle. The method of
center vortex ``removal'' subtracts the ideal center vortices arising in
the center projection from the fat vortices. This will give rise to
even larger gradients and accordingly the Dirac spectrum
Fig.~\ref{fig_000}(c) will show more ``noise.'' The method of center
vortex removal obviously works the better the thinner the vortices, and
becomes perfect for ideal center vortices (having a transversal
extension of one lattice spacing)\footnote{The method of vortex removal
converts an ideal center vortex into the 
trivial configuration $U_\mu = \Id$.}. 
The thinner the vortex the less flux
is left after vortex removal. To illustrate this we show in
Fig.~\ref{fig_001} the Dirac spectra for the vortex configurations which
result from those considered in Fig.~\ref{fig_000} when their
transversal vortex extension is shrunk to two lattice spacings. The
starting center vortices now have larger gradients and the Dirac
eigenvalues are now shifted off the Ginsparg-Wilson circle. At the same
time the Dirac eigenvalues of the vortex-removed configuration
Fig.~\ref{fig_001} (c) are now basically on the Ginsparg-Wilson circle
implying that this configuration contains only little gradients contrary
to what we have observed for the very fat vortices  (Fig.~\ref{fig_000}
(c)). For sake of completeness let us also mention that the center
projected configurations of the very fat and thin center vortices are
the same and so are the corresponding Dirac spectra shown in
Fig.~\ref{fig_000} (b) and Fig.~\ref{fig_001} (b), respectively.

\subsection{Eigenvectors}

After having demonstrated that for smooth center vortices the spectrum 
indeed matches the expectation from the index theorem, 
let us now analyze the corresponding eigenvectors.

\begin{figure}[p]
\begin{center}
\epsfig{file=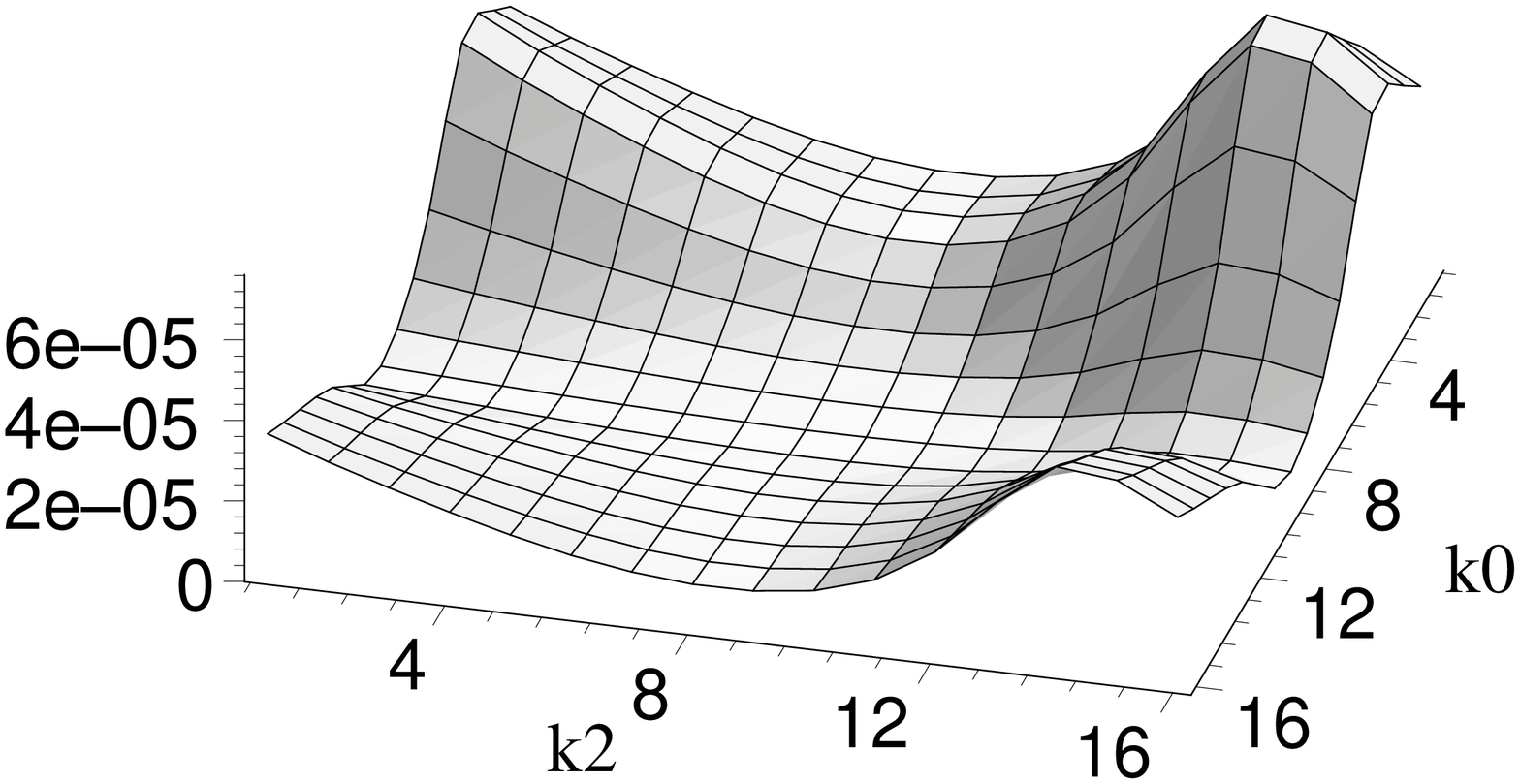,
        height=6cm,clip,angle=270}
\epsfig{file=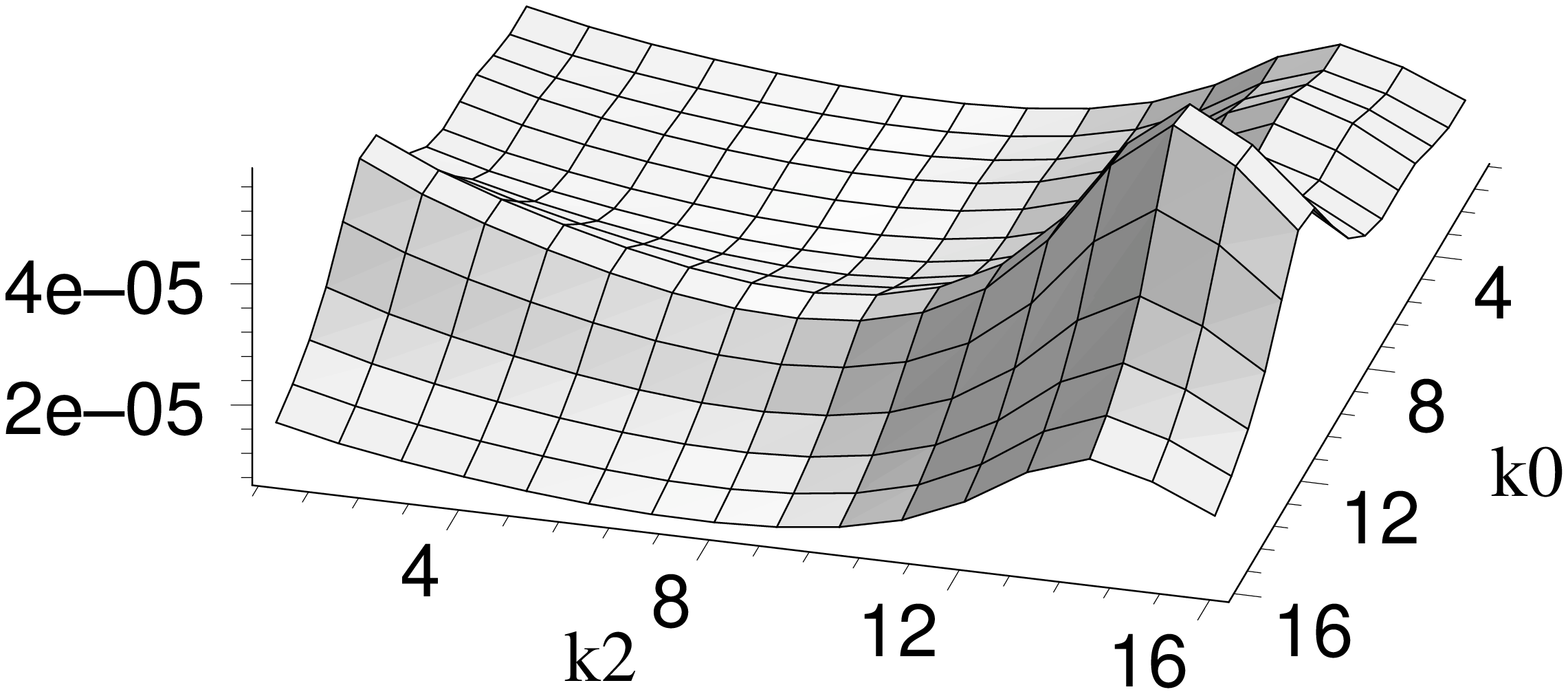,
        height=6cm,clip,angle=270}\\
\epsfig{file=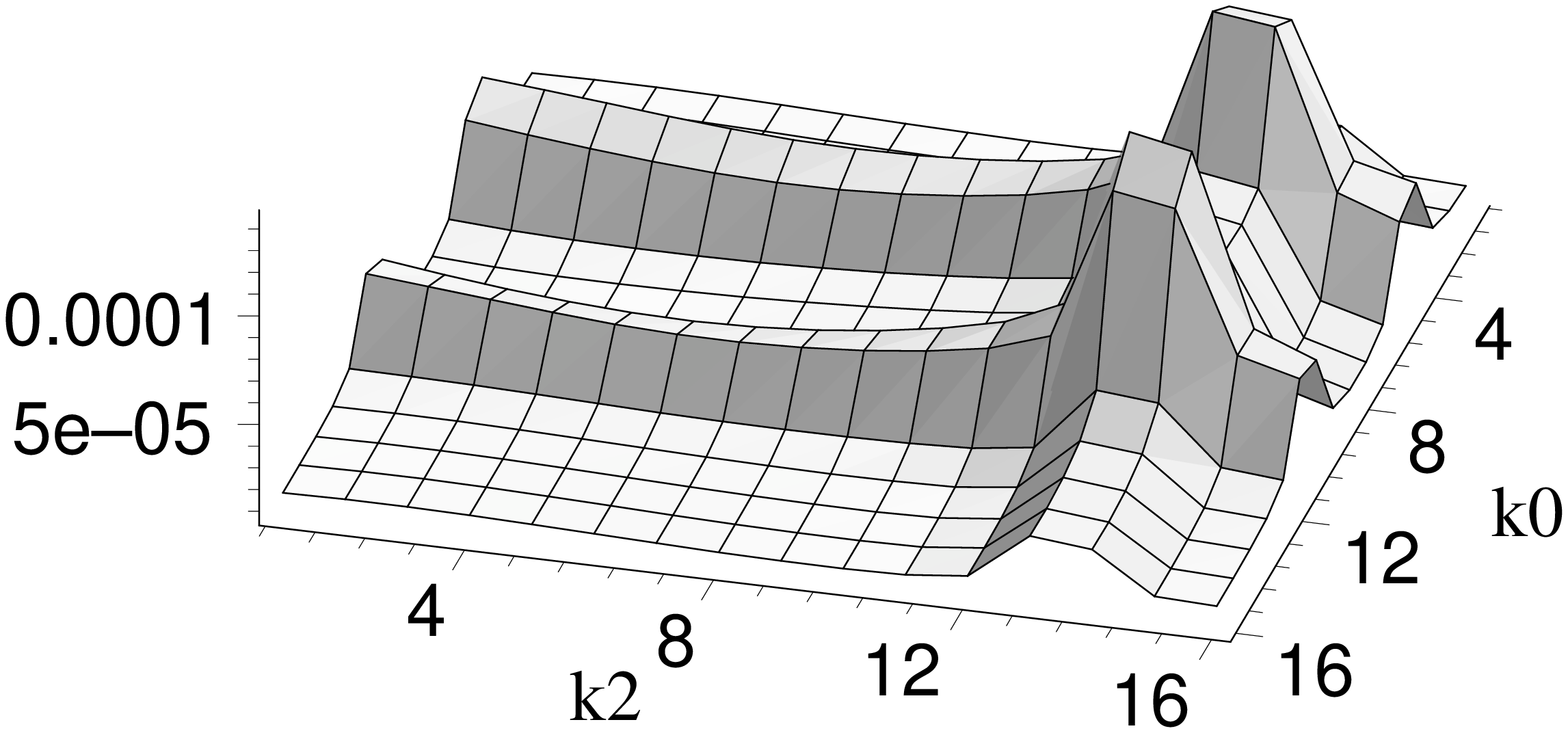,
        height=6cm,clip,angle=270}
\epsfig{file=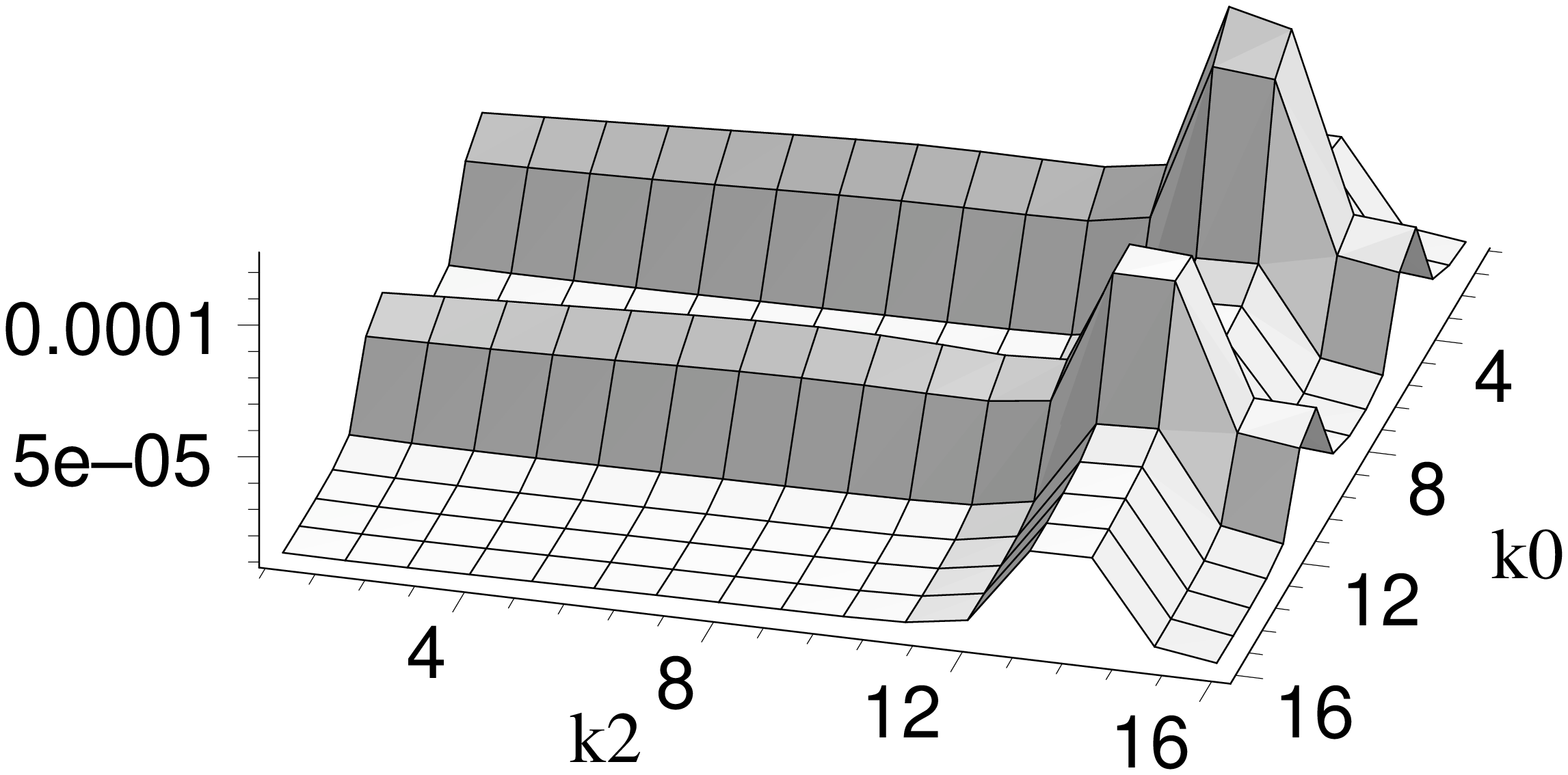,
        height=6cm,clip,angle=270}
\end{center}
\vspace{-5mm}
\caption{\label{fig_010}{\sl Plots of the scalar density of the lowest
non-zero eigenmodes of configurations $\# 0$ $($upper left$)$ and $\# 4$
$($upper right$)$ and of their center projected $($lower left$)$ and center
vortex-removed counter part $($lower right$)$ 
in the 2-d slice defined by $(k_1,k_3)=(12,14)$.
In this plane the configuration has intersection points at 
$(k_0, k_2) = (4,14)$ and $(10,14)$.}}
%
\begin{center}
\epsfig{file=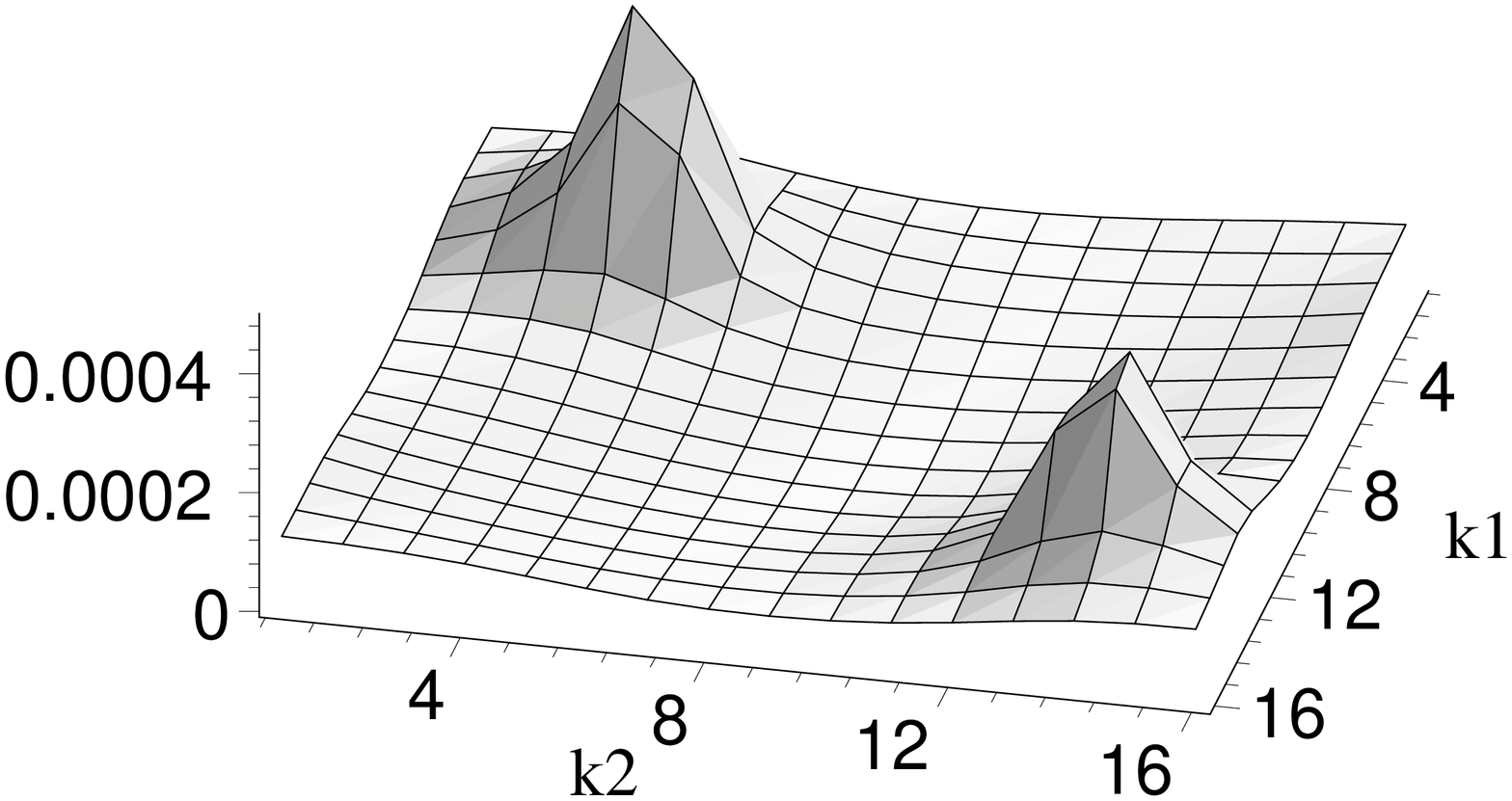,
        height=6cm,clip,angle=270} 
\epsfig{file=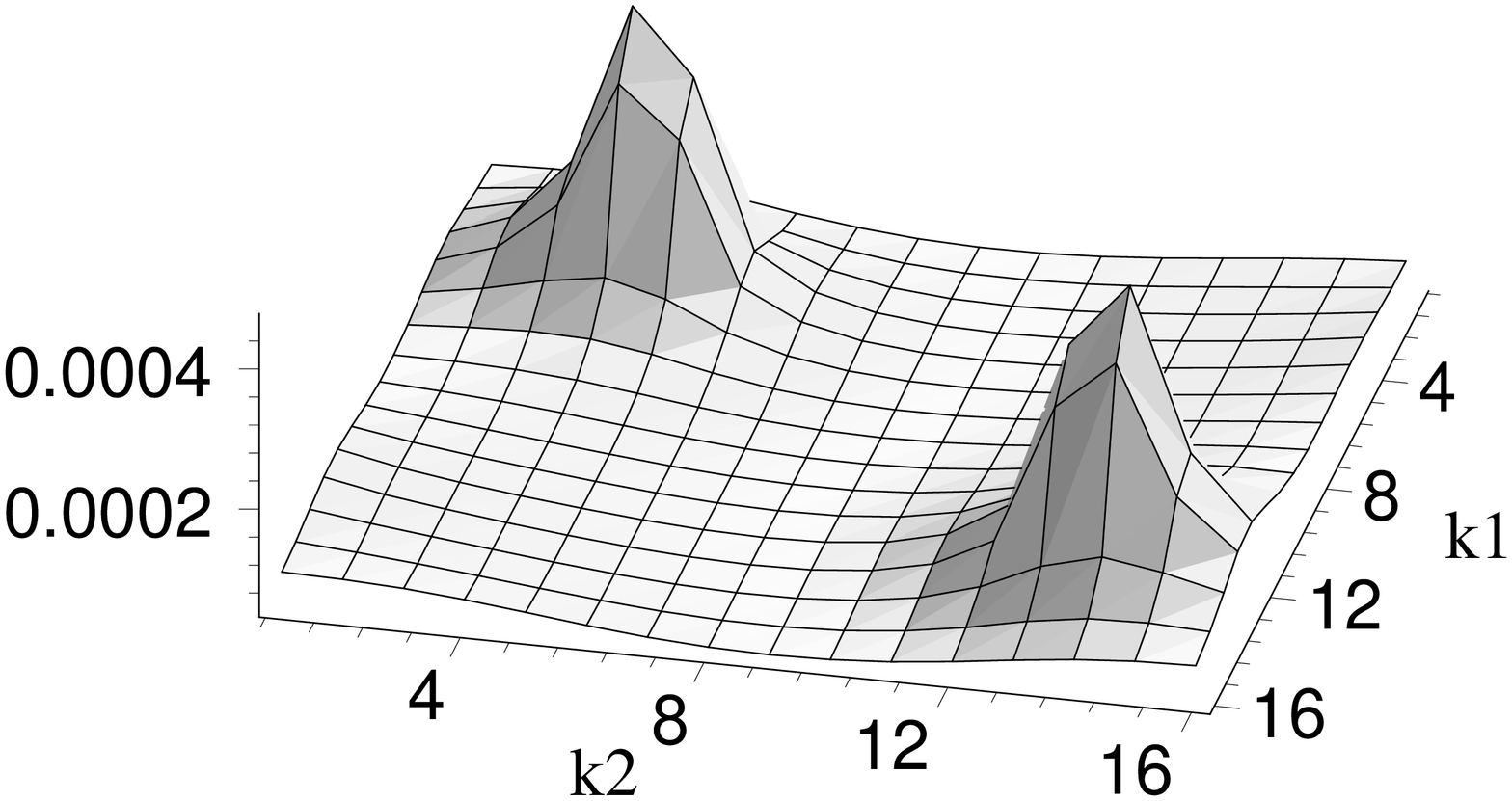,
        height=6cm,clip,angle=270} 
\\
\epsfig{file=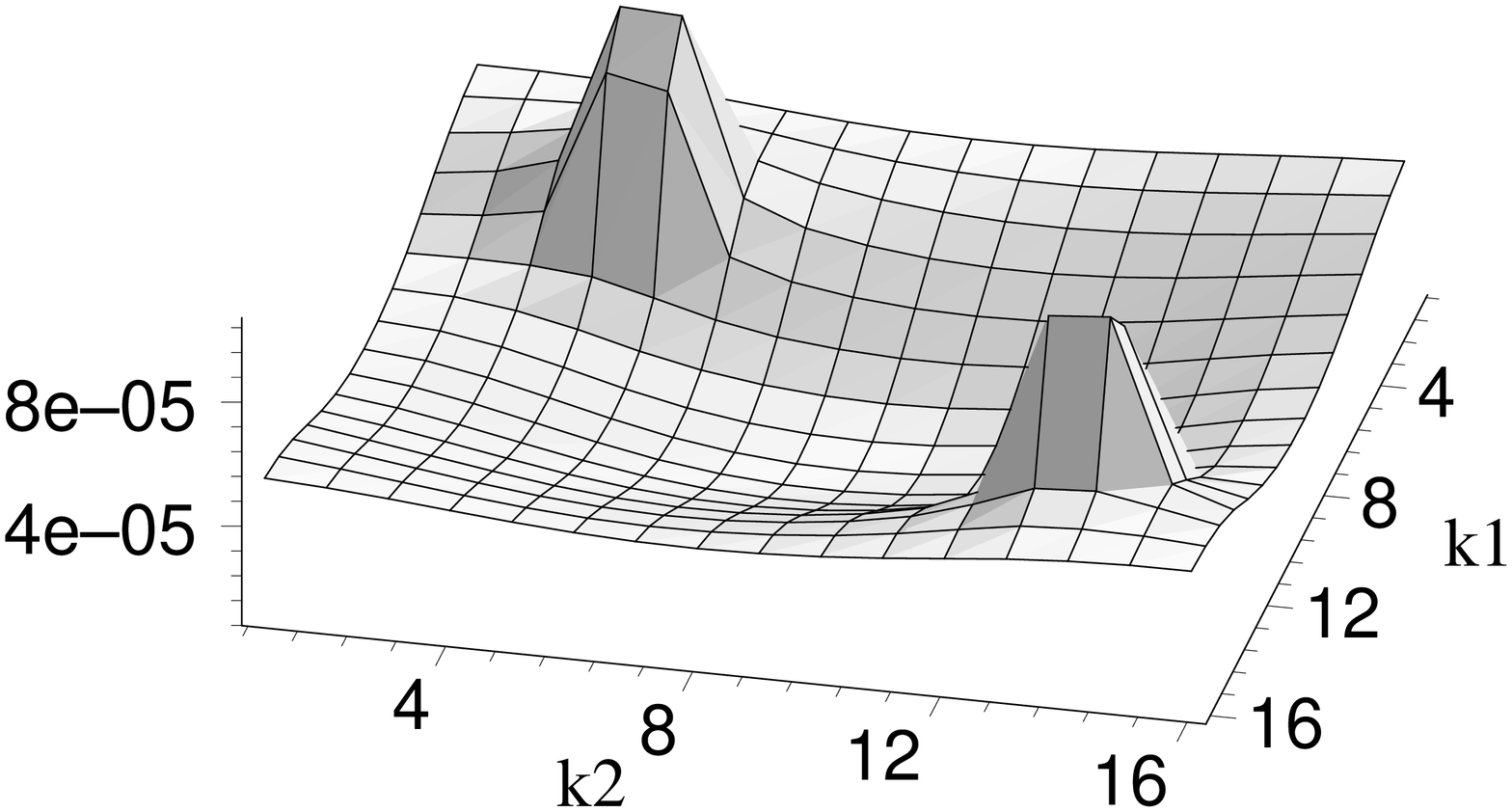,
        height=6cm,clip,angle=270} 
\epsfig{file=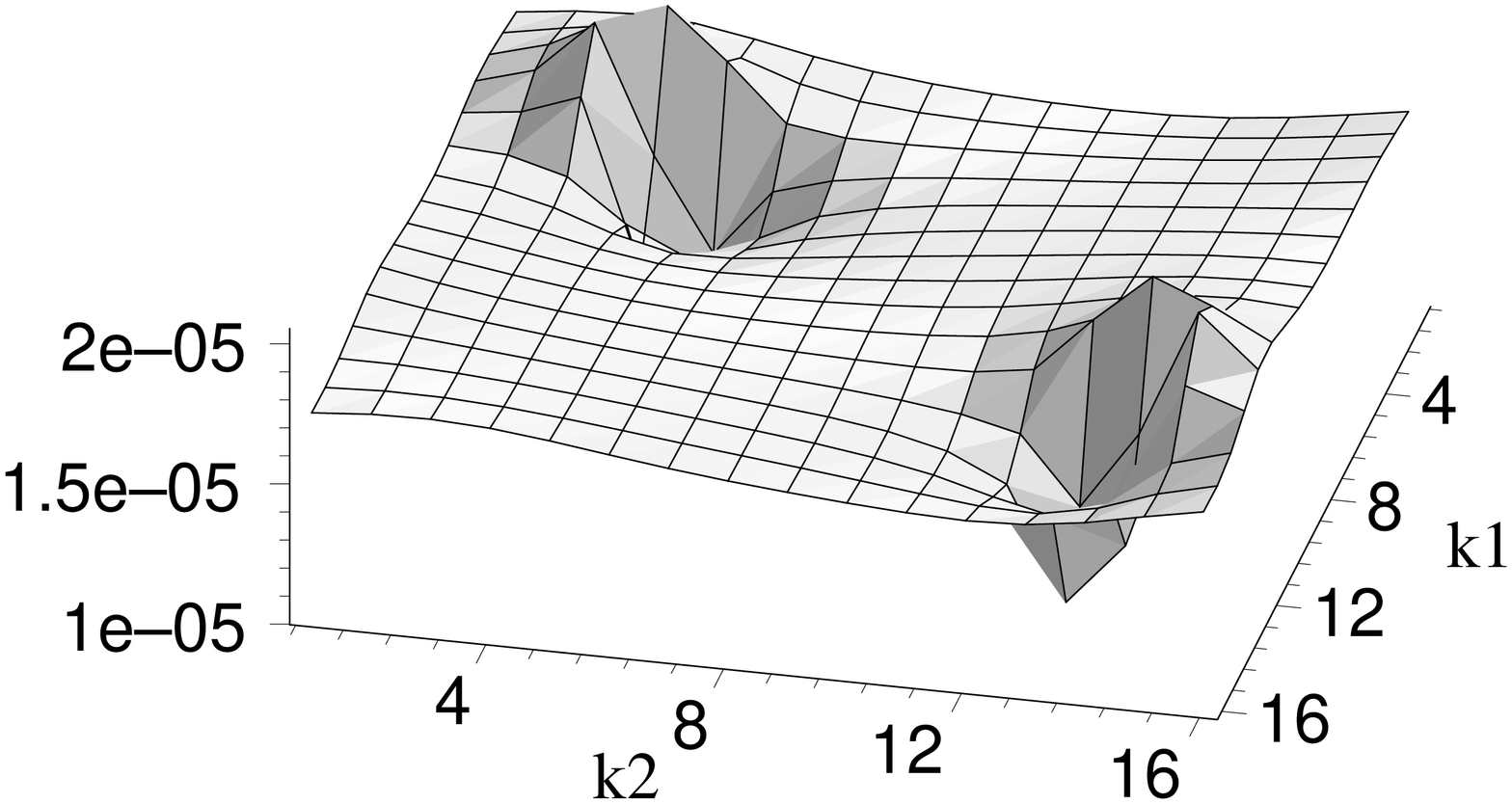,
        height=6cm,clip,angle=270} 
\end{center}
\vspace{-5mm}
\caption{\label{fig_009}{\sl Plots of the scalar density as in
Fig.~\ref{fig_010}, but in the 2-d slice defined by the second vortex, 
i.e.~$(k_0, k_3) = (4, 14)$, see Table \ref{tab_001}. In this plane 
the configuration has vortex intersection points at 
$(k_1, k_2) = (4, 4)$ and $(12, 14)$. Furthermore, 
we show the scalar density of the first non-zero mode for the 
corresponding vortex-removed configuration $($lower right$)$ and of the 
center projected configuration $($lower left$)$.}}
\end{figure}

Fig.~\ref{fig_010} shows the scalar density of the lowest lying non-zero
mode of the fat vortex configurations $\# 0$ ($\nu=2$) and $\# 4$ 
($\nu=0$) and their center projected image. Both configurations have up
to orientation the same same vortex flux and thus the same center
projected image. The scalar density is shown for a plane which does not 
coincide with one of the vortex planes. As is seen the scalar density 
of the lowest non-zero mode detects only one of the two parallel 
$k_1-k_2$ flux sheets at $k_0=4$ and $k_0=1$, and furthermore different 
sheets for the $\nu=2$ and $\nu=0$ configurations. The scalar density 
of the lowest Dirac mode of the center projected configuration sees both 
center vortex fluxes and in more pronounced form than the corresponding 
Dirac mode in the original fat center vortex background. Furthermore, 
the scalar density is concentrated at the vortex intersection points. 
These results are in accordance with the findings of 
Ref.~\cite{Reinhardt:2002cm} 
where the concentration of the scalar density of the zero modes near the
vortex flux and, in particular, at the intersection points was found. 
Surprisingly, the scalar density of the center vortex-removed
configuration is also localized at the vortex flux and is basically the
same as for the center projected configuration. One would have expected
that the localized structures arising from the center vortices and their
intersections are eliminated by the vortex removal. However, one should
keep in mind that vortex removal by the method of 
Ref.~\cite{deForcrand:1999ms},
implies subtraction of an ideal center vortex from a fat vortex in the 
original configuration. If the original vortex is very fat, vortex
removal near the vortex core
is almost the same as center projection. Only for thin smooth
center vortices the vortex removal procedure properly removes the vortex
flux. 

Fig.~\ref{fig_009} shows the scalar density for the same configurations
as in Fig.~\ref{fig_010}, however, for a different 2-d slice which 
coincides with a vortex plane. In addition, the transversal size of the
vortex flux was shrunk to two lattice spacings. Again, the scalar
density is localized at the intersection points and is basically the
same for original and the center projected configurations while it is
supressed for the vortex-removed configuration, in particular at
the intersection points (Note the different scales!).

Thus for very smooth fat
center vortex configurations the center projected vortex surfaces are
also smooth and their fluxes and topological spots are very well seen by
the scalar density of the near zero modes.

\begin{figure}[p]
\begin{center}
\begin{tabular}{lcccr}
& smooth vortex & center projected & vortex-removed \\
$\rho_+(x)$ & 
\epsfig{file=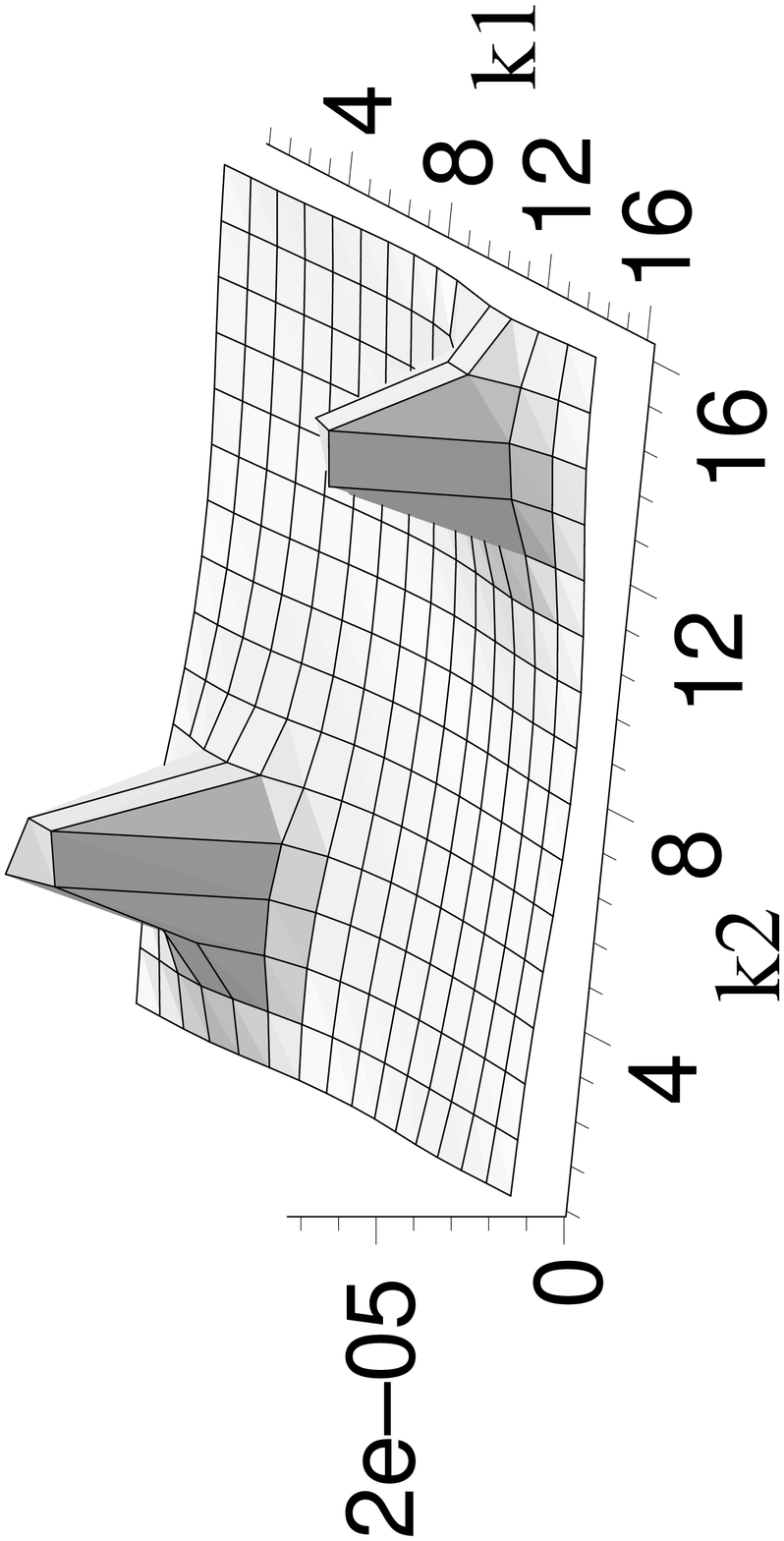,height=3.5cm,clip,angle=270} &
\epsfig{file=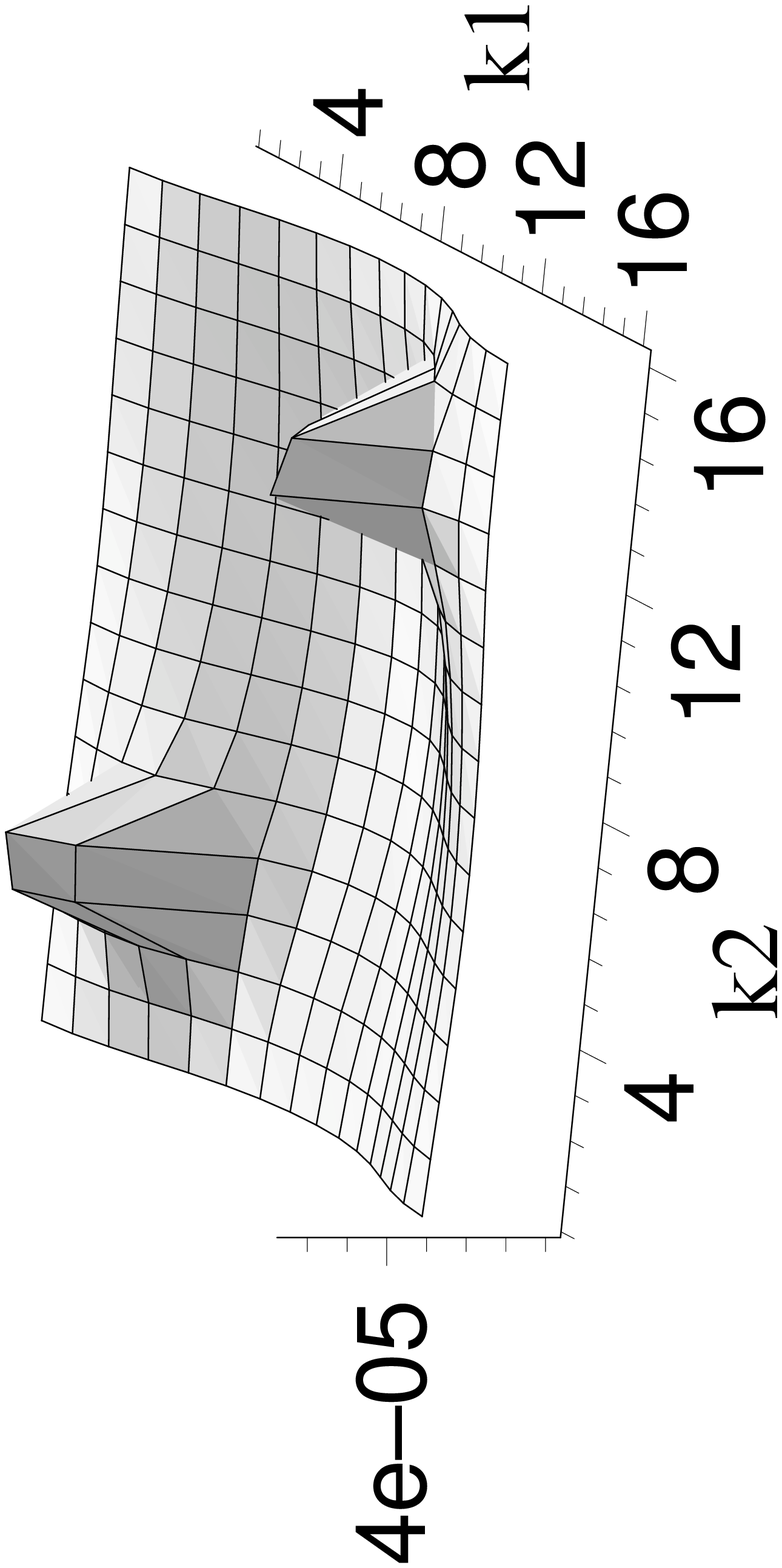,height=3.5cm,clip,angle=270} &
\epsfig{file=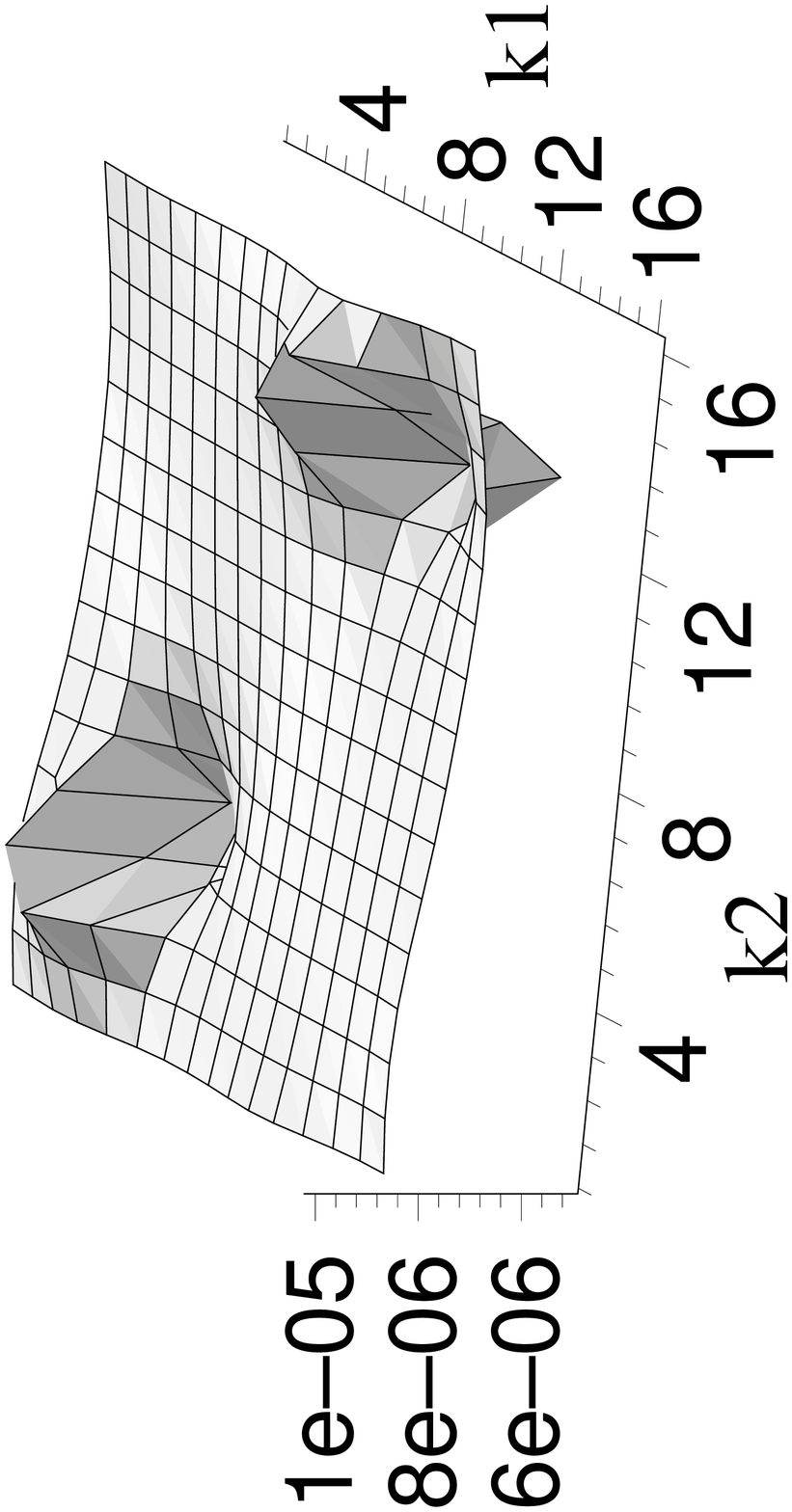,height=3.5cm,clip,angle=270} \\
$\rho_-(x)$ & 
\epsfig{file=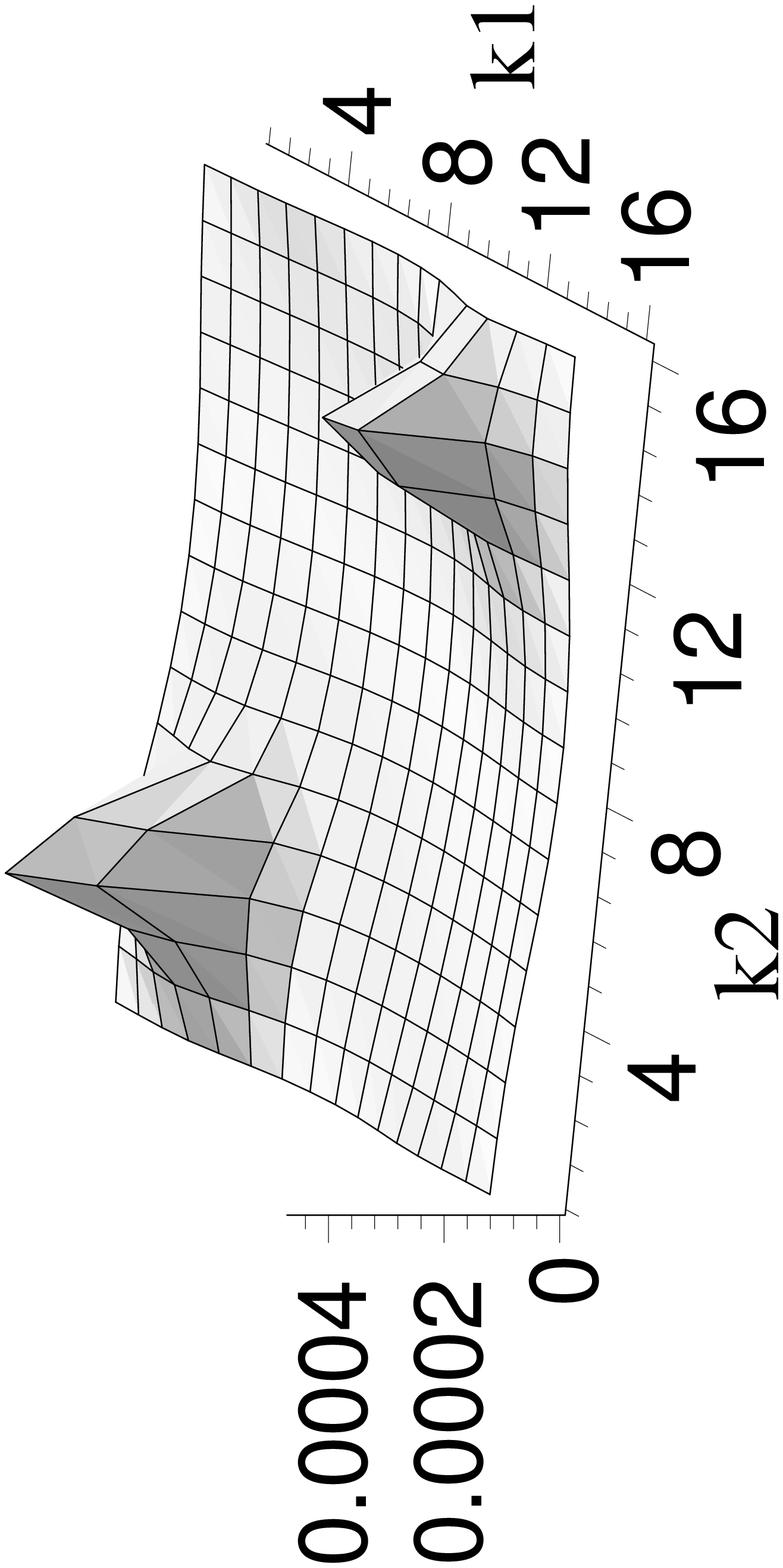,height=3.5cm,clip,angle=270} &
\epsfig{file=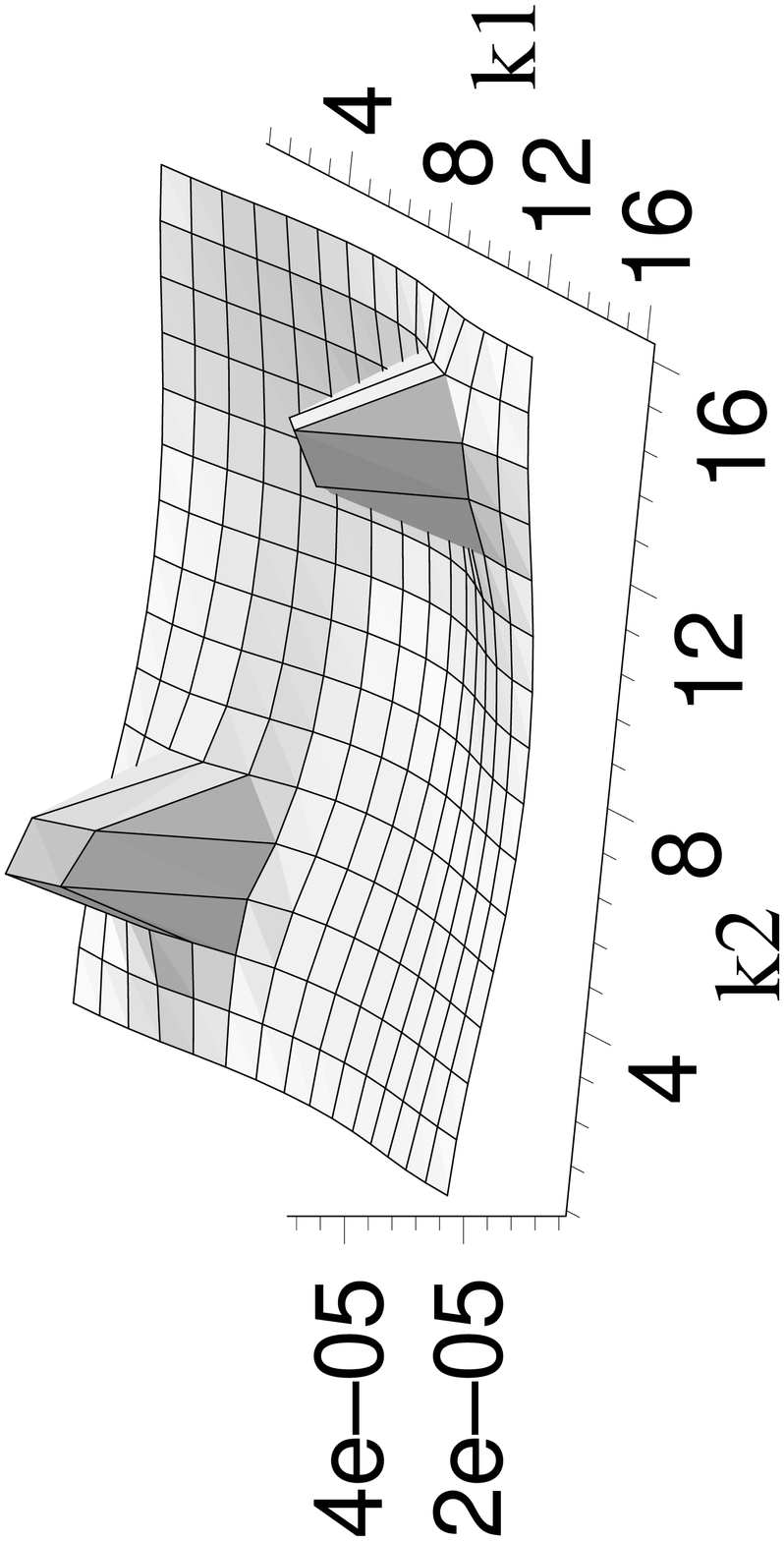,height=3.5cm,clip,angle=270} &
\epsfig{file=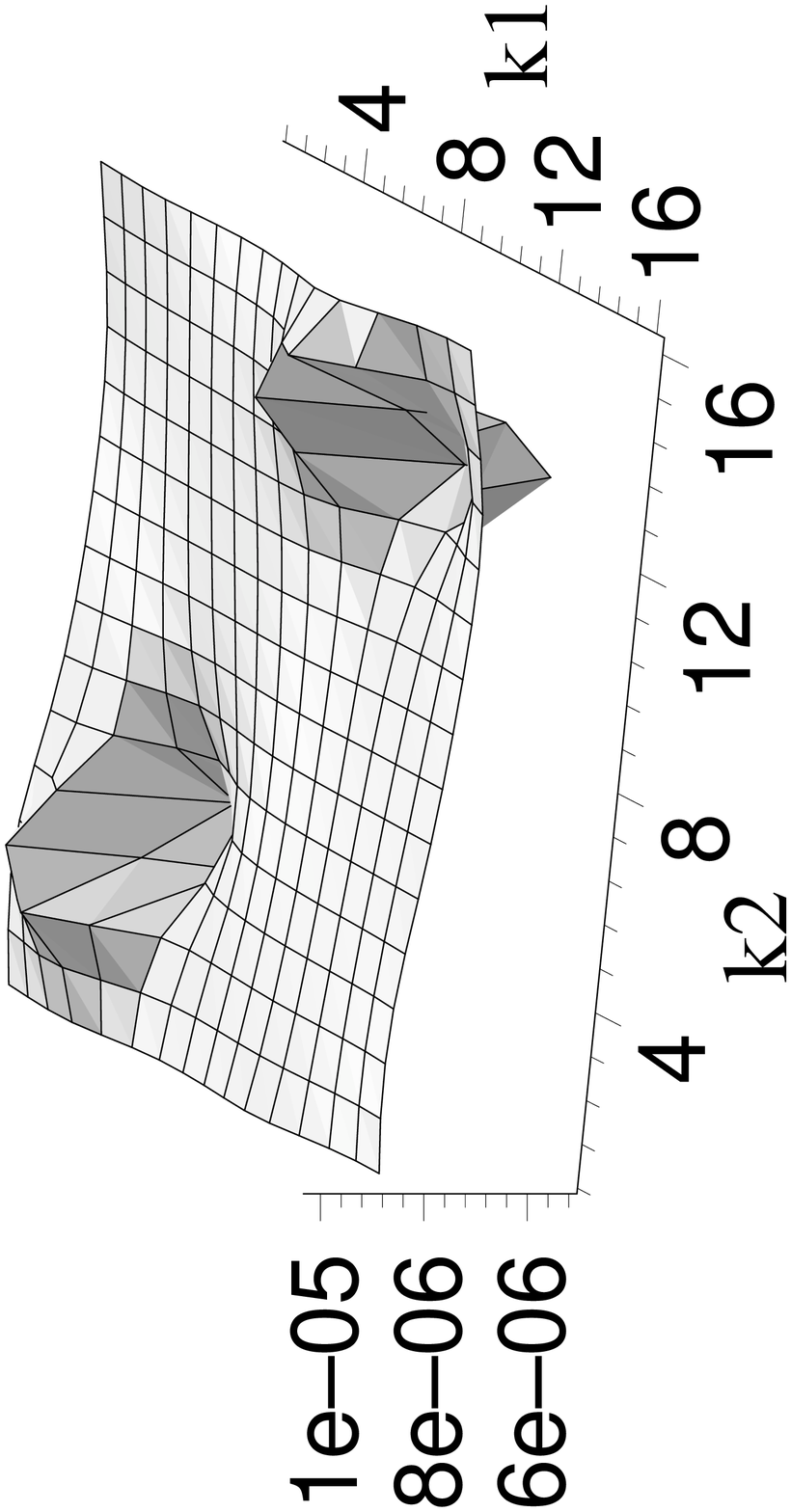,height=3.5cm,clip,angle=270} & (a)
\\ \\ \\
\hline
\\
\\
$\rho_+(x)$ & 
\epsfig{file=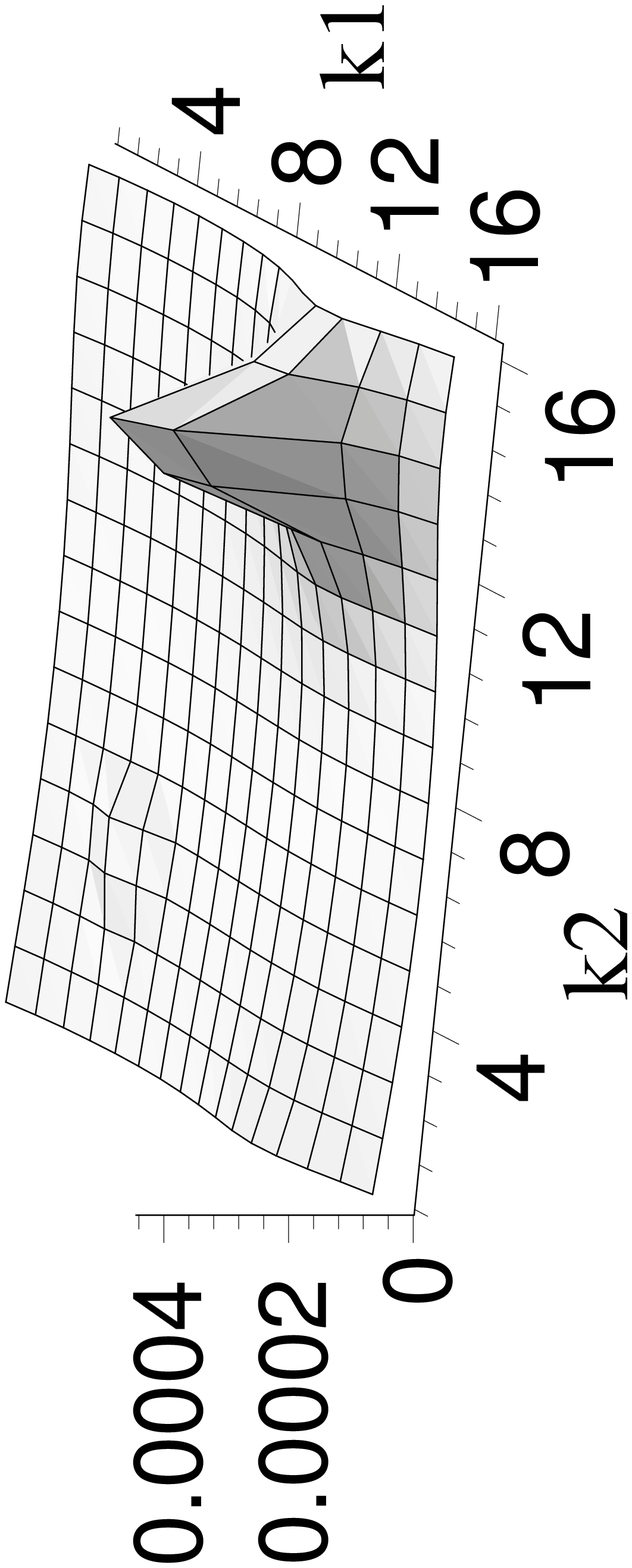,height=3.5cm,clip,angle=270} &
\epsfig{file=conf_112_rP.eps,height=3.5cm,clip,angle=270} &
\epsfig{file=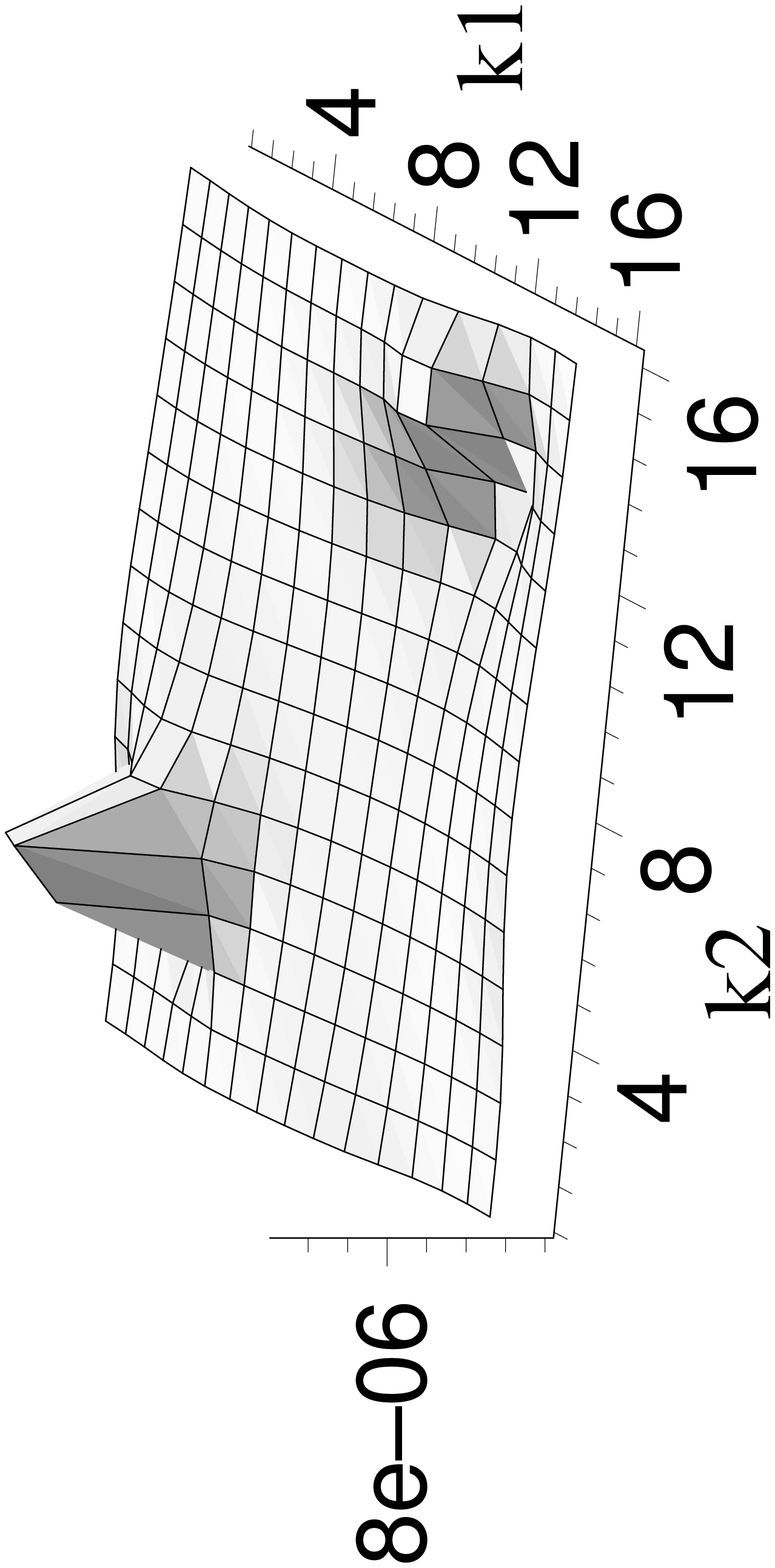,height=3.5cm,clip,angle=270} \\
$\rho_-(x)$ & 
\epsfig{file=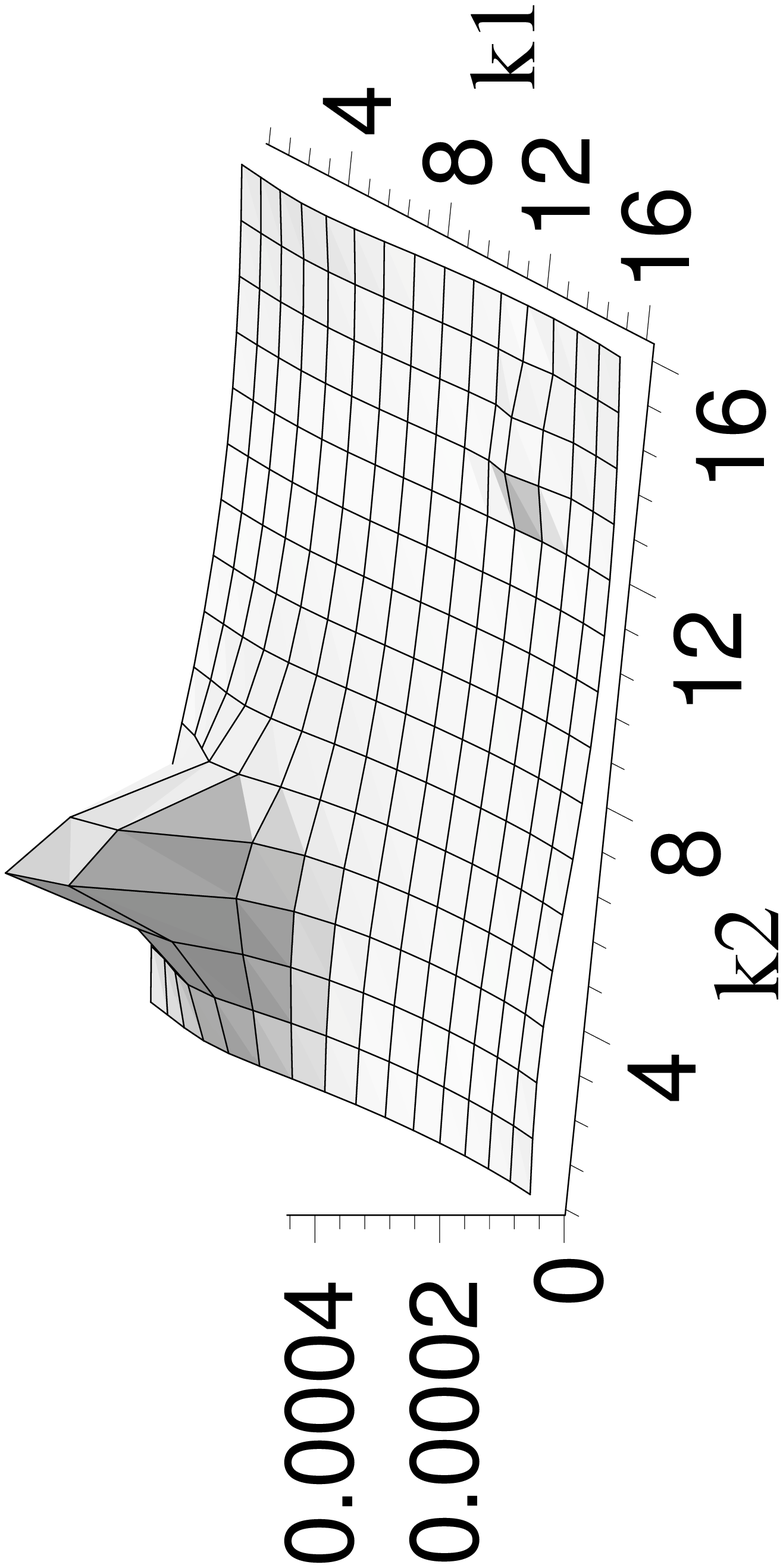,height=3.5cm,clip,angle=270} &
\epsfig{file=conf_112_rM.eps,height=3.5cm,clip,angle=270} &
\epsfig{file=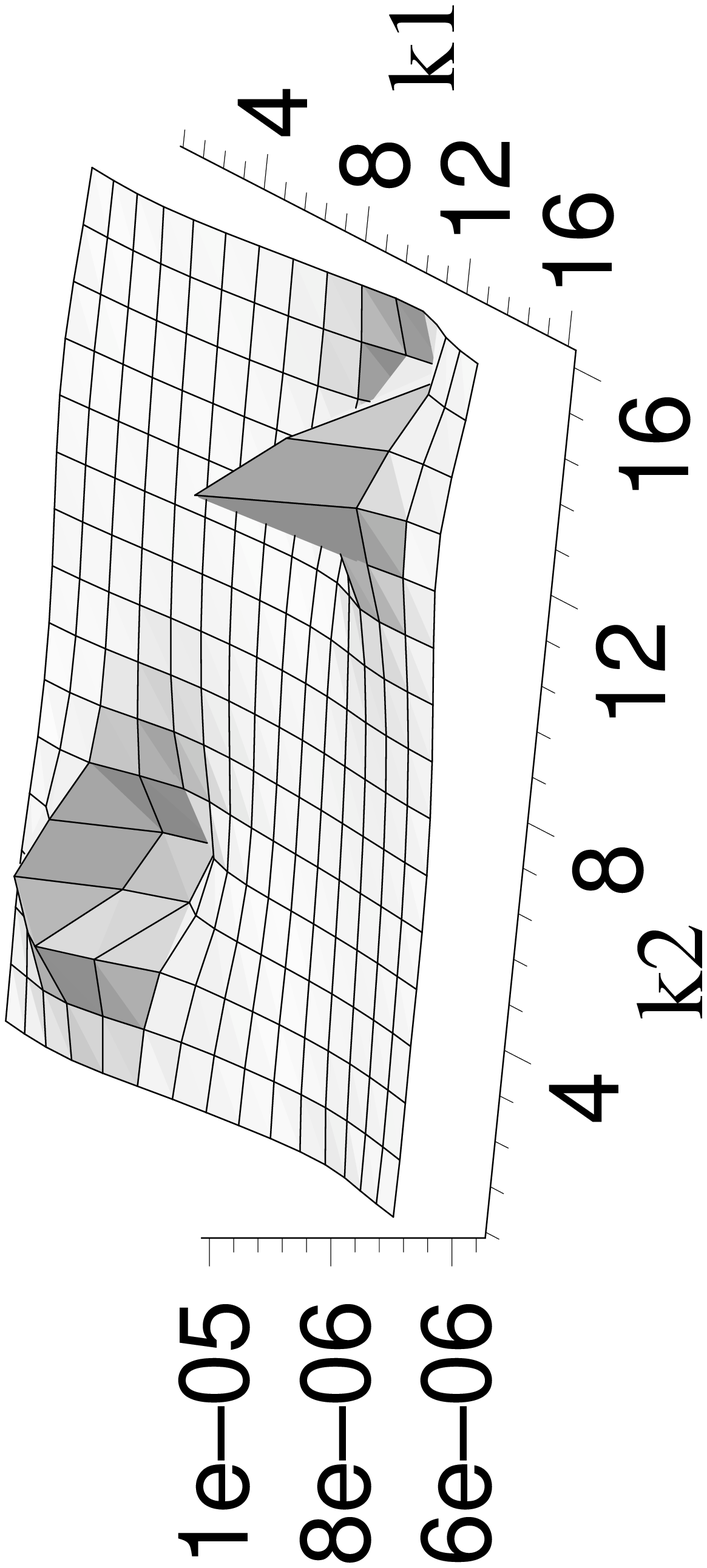,height=3.5cm,clip,angle=270} & (b)
\end{tabular}
\end{center}
\caption{\label{fig_013}
{\sl $($a$)$ Plots of the chiral densities $\rho_+(x)$ $($upper row$)$ and 
$\rho_-(x)$ $($lower row$)$ of the lowest non-zero Dirac mode for the
configuration $\# 0$ and its center projected and vortex-removed counter
parts through the 2-d slice defined by the plane $(k_0,k_3)=(4,14)$, in
which this configuration has two vortex intersection points. Note that
in the upper left plot the density $\rho_+(x)$ was stretched by a factor
of 20. $($b$)$ The same as in $($a$)$ for configuration $\# 4$.}}
\end{figure}

Contrary to the scalar density $\rho(x)$ (\ref{eq_005}), the chiral
densities $\rho_{\pm}(x)$ (\ref{eq_007}) do feel the orientation of the
flux. This can be seen in Fig.~\ref{fig_013}, where we show the chiral
densities of the lowest  non-zero Dirac modes of the configurations 
$\# 0$ and $\# 4$, which differ only in the orientation of the vortex
flux. The chiral densities are shown in a 2-d slice in which these two
configurations have two vortex intersection points.  
For configuration $\# 0$ the two vortex intersection points have the 
same orientation and accordingly give both rise to peaks of the same
size in the chiral densities $\rho_{\pm}(x)$. However, the density
$\rho_{+}(x)$ is suppressed by a factor of 20 compared to $\rho_{-}(x)$
implying that the considered non-zero Dirac mode is at both vortex
intersection points approximately left handed (negative chirality) in
agreement with the vortex orientation. For configuration
$\# 4$ the two intersection points have opposite orientation and 
we observe a peak in either $\rho_{+}(x)$ or $\rho_{-}(x)$ at the
two intersection points. The orientation of the flux is lost in the 
center projection which hence converts both center vortex 
configurations $\# 0$ and $\# 4$ into the same ideal center vortex. 
Therefore the center projected configurations do not distinguish between
$\rho_{+}(x)$ and $\rho_{-}(x)$ and consequently do not 
reproduce the chiral densities properly. This is indeed seen in 
Fig.~\ref{fig_013} where we compare the chiral
densities of the lowest lying non-zero Dirac modes for the fat center 
vortex configurations $\# 0$ and $\# 4$, the corresponding center 
projected and the center vortex-removed configurations. 
%


\section{Summary and Conclusions}

Using the chirally improved Dirac operator constructed 
in~\cite{Gattringer:2000js,Gattringer:2000qu}, 
we have studied the influence of center vortices on various properties
of the Dirac spectrum. We have shown that removal of center vortices
eliminates the zero modes and near zero modes of the Dirac operator. 
Via the Banks-Casher relation this implies the restoration of chiral symmetry. 
We have demonstrated 
that the spectrum for vortex-removed configurations strongly
resembles the free Dirac spectrum. For the Dirac eigenmodes we found that 
removing the center vortices destroys local chirality of the low
lying modes. Comparison with configurations, subject 
to incoherent random changes, 
shows that removing the center vortices specifically targets the topological
properties, while random changes leave them intact. 

Somewhat surprisingly we find that also center projection eliminates 
the topological charge and gives rise to a gap near zero virtuality. 
However, this result
does not mean that center vortices are not responsible for the
topological properties of the gauge fields and for the spontaneous
breaking of chiral symmetry but can be traced back to the fact that
center projection removes the orientation of the vortices. Center
projection keeps track, however, of the module of the flux of the
vortices. This was demonstrated by measuring the scalar density of the
lowest-lying non-zero modes, which is insensitive to the direction of
the flux. The center projected configurations show the same localization
of the scalar density near the center vortices (and their intersection
points) as the full configurations when the latter are rather smooth. 
While the center projected 
configurations reproduce the localization of the scalar density near the
vortices, they do not reproduce the localization of the chiral densities,
as the latter depend on the orientation of the flux. The orientation of
the flux of the center vortices is irrelevant for their confining
properties but crucial for their topological and chiral properties.
Therefore we arive at the conclusion that the familiar method of center
projection is not suited to study the chiral and topological properties
of center vortices for lattice ensembles. To capture the topological 
properties one has, at least, to embed the center configurations 
into the Cartan subgroup to preserve the orientation of the 
center surfaces.


\section*{Acknowledgments:} 

We thank Manfried Faber, Meinulf G\"ockeler, Christian Lang,
Rainer Pullirsch and Pierre van Baal for discussions and the referee 
for remarks that were included in the last paragraph of Section 2.3. 
The calculations were done on the Hitachi SR8000
at the Leibniz Rechenzentrum in Munich and we thank the LRZ staff for
training and support. This work was supported by the DFG Forschergruppe
``Lattice-Hadron-Phenomenology''and by DFG - Re856/5-1.  



\begin{thebibliography}{1234567}


\bibitem{DelDebbio:1996mh} L.~Del Debbio, M.~Faber, J.~Greensite and 
 S.~Olejnik,
 Phys.\ Rev.\ {\bf D55}, 2298 (1997)

\bibitem{Langfeld:1997jx}
K.~Langfeld, H.~Reinhardt and O.~Tennert,
Phys.\ Lett.\ B {\bf 419}, 317 (1998)

\bibitem{deForcrand:1999ms}
P.~de Forcrand and M.~D'Elia,
Phys.\ Rev.\ Lett.\  {\bf 82}, 4582 (1999)

\bibitem{Alexandrou:1999vx}
C.~Alexandrou, P.~de Forcrand and M.~D'Elia,
Nucl.\ Phys.\ A {\bf 663}, 1031 (2000)

\bibitem{Langfeld:2001cz}
K.~Langfeld, H.~Reinhardt and J.~Gattnar,
Nucl.\ Phys.\ B {\bf 621}, 131 (2002)

\bibitem{Gattnar:2004bf}
J.~Gattnar, K.~Langfeld and H.~Reinhardt,
Phys.\ Rev.\ Lett.\  {\bf 93}, 061601 (2004)


\bibitem{Greensite:2004ur}
J.~Greensite, S.~Olejnik and D.~Zwanziger,
arXiv:hep-lat/0407032

\bibitem{Langfeld:1998cz}
K.~Langfeld, O.~Tennert, M.~Engelhardt and H.~Reinhardt,
Phys.\ Lett.\ B {\bf 452}, 301 (1999)

\bibitem{Engelhardt:1999fd}
M.~Engelhardt, K.~Langfeld, H.~Reinhardt and O.~Tennert,
Phys.\ Rev.\ D {\bf 61}, 054504 (2000)

\bibitem{Langfeld:2003zi}
K.~Langfeld,
Phys.\ Rev.\ D {\bf 67}, 111501 (2003) 
 
\bibitem{Gattnar:2000ke}
J.~Gattnar, K.~Langfeld, A.~Schafke and H.~Reinhardt,
Phys.\ Lett.\ B {\bf 489}, 251 (2000) 

\bibitem{Engelhardt:1999wr}
M.~Engelhardt and H.~Reinhardt,
Nucl.\ Phys.\ B {\bf 585}, 591 (2000)

\bibitem{Engelhardt:2003wm}
M.~Engelhardt, M.~Quandt and H.~Reinhardt,
Nucl.\ Phys.\ B {\bf 685}, 227 (2004)

\bibitem{Kovacs:2000sy}
T.~G.~Kovacs and E.~T.~Tomboulis,
Phys.\ Rev.\ Lett.\  {\bf 85}, 704 (2000)

\bibitem{Lange:2003ti}
J.~D.~L\"ange, M.~Engelhardt and H.~Reinhardt,
Phys.\ Rev.\ D {\bf 68}, 025001 (2003)

\bibitem{Reinhardt:2002cm}
H.~Reinhardt, O.~Schroeder, T.~Tok and V.~C.~Zhukovsky,
Phys.\ Rev.\ D {\bf 66}, 085004 (2002)

\bibitem{DelDebbio:1998uu} L.~Del Debbio, M.~Faber, J.~Giedt, 
 J.~Greensite and S.~Olejnik,
 Phys.\ Rev.\ {\bf D58}, 094501 (1998)

\bibitem{Engelhardt:1999xw}
M.~Engelhardt and H.~Reinhardt,
Nucl.\ Phys.\ B {\bf 567}, 249 (2000)

\bibitem{Reinhardt:2001kf}
H.~Reinhardt,
Nucl.\ Phys.\ B {\bf 628}, 133 (2002)

\bibitem{Gattringer:2000js}
C.~Gattringer,
Phys.\ Rev.\ D {\bf 63}, 114501 (2001)

\bibitem{Gattringer:2000qu}
C.~Gattringer, I.~Hip and C.~B.~Lang,
Nucl.\ Phys.\ B {\bf 597}, 451 (2001)

\bibitem{Gattringer:2001cf}
C.~Gattringer, M.~G\"ockeler, C.~B.~Lang, P.~E.~L.~Rakow and A.~Sch\"afer,
Phys.\ Lett.\ B {\bf 522}, 194 (2001)

\bibitem{arnoldi} D.C.~Sorensen, SIAM J.~Matrix, 
 Anal.\ Appl.\ {\bf 13}, 357 (1992) 

\bibitem{Gattringer:2002wh}
C.~Gattringer,
Phys.\ Rev.\ D {\bf 67}, 034507 (2003)

\bibitem{Gattringer:2002tg}
C.~Gattringer and S.~Schaefer,
Nucl.\ Phys.\ B {\bf 654}, 30 (2003)

\bibitem{Gattringer:2004pb}
C.~Gattringer and R.~Pullirsch,
Phys.\ Rev.\ D {\bf 69}, 094510 (2004)

\bibitem{Gattringer:2004va}
C.~Gattringer and S.~Solbrig,
arXiv:hep-lat/0410040

\bibitem{Banks:1979yr}
T.~Banks and A.~Casher,
Nucl.\ Phys.\ B {\bf 169}, 103 (1980)

\bibitem{Gattringer:1997ci}
C.~Gattringer and I.~Hip,
Nucl.\ Phys.\ B {\bf 536}, 363 (1998)


\bibitem{Horvath:2003yj}
I.~Horvath {\it et al.},
Phys.\ Rev.\ D {\bf 68}, 114505 (2003)

\bibitem{Horvath:2002yn}
I.~Horvath, S.~J.~Dong, T.~Draper, F.~X.~Lee, K.~F.~Liu, H.~B.~Thacker and J.~B.~Zhang,
Phys.\ Rev.\ D {\bf 67}, 011501 (2003)

\bibitem{Aubin:2004mp}
C.~Aubin {\it et al.}  [MILC Collaboration],
arXiv:hep-lat/0410024

\bibitem{Horvath:2001ir}
I.~Horvath, N.~Isgur, J.~McCune and H.~B.~Thacker,
Phys.\ Rev.\ D {\bf 65}, 014502 (2002)

\bibitem{Horvath:2002gk}
I.~Horvath {\it et al.},
Phys.\ Rev.\ D {\bf 66}, 034501 (2002)

\bibitem{DeGrand:2001pj}
T.~DeGrand and A.~Hasenfratz,
Phys.\ Rev.\ D {\bf 65}, 014503 (2002)

\bibitem{Blum:2001qg}
T.~Blum {\it et al.},
Phys.\ Rev.\ D {\bf 65}, 014504 (2002)

\bibitem{Edwards:2001nd}
R.~G.~Edwards and U.~M.~Heller,
Phys.\ Rev.\ D {\bf 65}, 014505 (2002)

\bibitem{Hip:2001hc}
I.~Hip, T.~Lippert, H.~Neff, K.~Schilling and W.~Schroers,
Phys.\ Rev.\ D {\bf 65}, 014506 (2002)

\bibitem{Gattringer:2001mn}
C.~Gattringer, M.~G\"ockeler, P.~E.~L.~Rakow, S.~Schaefer and A.~Sch\"afer,
Nucl.\ Phys.\ B {\bf 617}, 101 (2001)

\bibitem{Gattringer:2001ia}
C.~Gattringer, M.~G\"ockeler, P.~E.~L.~Rakow, S.~Schaefer and A.~Sch\"afer,
Nucl.\ Phys.\ B {\bf 618}, 205 (2001)

\bibitem{Hasenfratz:2001qp}
P.~Hasenfratz, S.~Hauswirth, K.~Holland, T.~J\"org and F.~Niedermayer,
Nucl.\ Phys.\ Proc.\ Suppl.\  {\bf 106}, 751 (2002)

\end{thebibliography}
\end{document}